\documentclass[12pt,a4paper,oneside]{article}
\usepackage{ifpdf}
\usepackage{a4wide}
\usepackage{amsmath}
\usepackage{graphicx}
\usepackage{rotating}
\usepackage[numbers,sort&compress]{natbib} 
\usepackage{amssymb}
\begin{document}

\textheight 21.0cm
\textwidth 16cm
\sloppy
\oddsidemargin 0.0cm \evensidemargin 0.0cm
\topmargin 0.0cm

\setlength{\parskip}{0.45cm}
\setlength{\baselineskip}{0.75cm}



\begin{titlepage}
\setlength{\parskip}{0.25cm}
\setlength{\baselineskip}{0.25cm}
\begin{flushright}
DO-TH 08/05\\
\vspace{0.2cm}
September 2008
\end{flushright}
\vspace{1.0cm}
\begin{center}
\Large
{\bf Dynamical NNLO parton distributions}
\vspace{1.5cm}

\large
P.~Jimenez-Delgado, E.\ Reya
\vspace{1.0cm}

\normalsize
{\it Universit\"{a}t Dortmund, Institut f\"{u}r Physik}\\
{\it D-44221 Dortmund, Germany} \\

\vspace{1.5cm}
\end{center}

\begin{abstract}
\noindent 
Utilizing recent DIS measurements ($\sigma_r, F_{2,3,L}$) and data
on hadronic dilepton production we determine at NNLO (3-loop) of 
QCD the dynamical parton distributions of the nucleon generated
radiatively from valencelike {\em positive} input distributions at
an optimally chosen low resolution scale ($Q^2_0<1$ GeV$^2$).  
These are compared with `standard' NNLO distributions generated 
from positive input distributions at some {\em fixed} and higher 
resolution scale ($Q_0^2> 1$ GeV$^2$).  
Although the NNLO corrections imply in both approaches an improved
value of $\chi^2$, 
typically $\chi_{\rm NNLO}^2\simeq 0.9\chi^2_{\rm NLO}$, present DIS
data are still not sufficiently accurate to distinguish between
NLO results and the minute NNLO effects of a few percent, despite
of the fact that the dynamical NNLO uncertainties are somewhat
smaller than the NLO ones and both are, as expected, smaller than
those of their `standard' counterparts.  The dynamical predictions
for $F_L(x,Q^2)$ become perturbatively stable already at $Q^2=2-3$
GeV$^2$ where precision measurements could even delineate NNLO
effects in the very small--$x$ region.  This is in contrast to the
common  `standard' approach but NNLO/NLO differences are here less
distinguishable due to the larger $1\sigma$ uncertainty bands.
Within the dynamical approach we obtain $\alpha_s(M_Z^2)=0.1124\pm
0.0020$, whereas the somewhat less constrained `standard' fit
gives $\alpha_s(M_Z^2)=0.1158\pm 0.0035$.  
\end{abstract}
\end{titlepage}

\section{Introduction}
Within the dynamical parton model approach the predicted small 
Bjorken-$x$ behavior of structure functions is entirely due to QCD 
dynamics at $x$ \raisebox{-0.1cm}{$\stackrel{<}{\sim}$} $10^{-2}$. 
This is due to the fact that the parton distributions at 
{$Q^2$~\raisebox{-0.1cm}{$\stackrel{>}{\sim}$} 1 GeV$^2$ are QCD
radiatively generated from {\em{valencelike}} positive input
distributions at an optimally determined low input scale 
$Q_0^2\equiv \mu^2<$ 1 GeV$^2$ (where `valencelike' refers to 
$a_f>0$ for {\em{all}} input distributions $xf(x,\mu^2)\propto 
x^{a_f}(1-x)^{b_f}$, i.e., not only the valence but also the 
sea and gluon input densities vanish at small $x$). Originally, 
its characteristic unique steep
small-$x$ predictions for the experimentally then unexplored region
$x<10^{-2}$ \cite{ref1,ref2,ref3} were subsequently first confirmed
in \cite{ref4,ref5}.  With the advent of further high--precision
data in recent years, the original dynamical parton distributions
had to be updated \cite{ref6,ref7,ref8} but the characteristic steep
small-$x$ behavior of the sea and the gluon distributions as $x\to 0$
remained essentially very similar.

Alternatively, in the common `standard' approach, e.g.\ 
[9--21], the input scale is fixed at some
{\em{arbitrarily}} chosen $Q_0>1$ GeV and the corresponding input
distributions are less restricted.  For example, the observed steep
small-$x$ behavior ($a_f<0$) of the gluon and sea distributions has to 
be {\em{fitted}}, allowing even for negative gluon distributions
\cite{ref12,ref13,ref14,ref18,ref19}, i.e.\ negative cross sections 
like a negative longitudinal structure function $F_L(x,Q^2)$.
Furthermore the associated uncertainties encountered in the determination
of the parton distributions turn out to be larger, particularly in the 
small-$x$ region, than in the more restricted dynamical radiative
approach where, moreover, the evolution distance (starting at 
$Q_0^2<1$ GeV$^2$) is sizeably larger.

In the present paper we extend our most recent LO and NLO dynamical
analysis \cite{ref8} to the next--to--next--to--leading order (NNLO)
of QCD.  For consistency reasons we only consider deep inelastic
scattering (DIS) and Drell--Yan dimuon production data where all 
required theoretical NNLO ingredients are available by now,
except the ones for heavy quark production. High-$p_T$ hadron--hadron
scattering processes will now not be considered since so far they are
only known up to NLO.  Furthermore we compare these  `dynamical' 
results for the radiatively generated parton distributions arising
from a valencelike positive input at $Q_0< 1$ GeV with the ones 
obtained from the common NNLO evolution approach being based on a
`standard' non--valencelike input at $Q_0>1$ GeV.  In addition we shall
analyze their associated uncertainties.  Section 2 will be devoted to
a discussion of some theoretical issues relevant for our NNLO analysis, 
in particular concerning our NNLO $Q^2$-evolution algorithm for parton
distributions, as well as of the relevant couplings and coefficient
functions for relating them to the various neutral current structure
functions required for calculating `reduced' DIS cross sections. 
In Sect.~3 we present our quantitative results for structure functions,
in particular our dynamical small-$x$ predictions,
and for hadronic dilepton production.  The results related to the 
longitudinal structure $F_L(x,Q^2)$ are discussed in Sect.~4.
Furthermore these
dynamical results, together with their associated $1\sigma$ uncertainties,
are compared with the ones obtained from a common `standard'
approach.
Our conclusions are summarized in Sect.~5.

\section{Evolutions of parton distributions and structure functions}

Our NNLO analyses will be performed within the modified minimal 
subtraction ($\overline{\rm MS}$) factorization and renormalization
scheme.  Heavy quarks ($c,b,t$) will not be considered as massless
partons within the nucleon, i.e.\ the number of active (light) flavors
$n_f$ appearing in the splitting functions and the corresponding 
Wilson coefficients will be fixed, $n_f=3$.  This defines the so--called
`fixed flavor number scheme' (FFNS) which is fully predictive in the 
heavy quark sector where the heavy quark flavors are produced entirely
perturbatively from the initial light ($u,d,s$) quarks and gluons -- in
full agreement with present experiments.  Furthermore, in the evaluation
of the running strong coupling $\alpha_s(Q^2)$ it is nevertheless 
consistent and correct to utilize the standard variable $n_f$ scheme
for the $\beta$-function \cite{ref22}. Up to NNLO, $a_s(Q^2)
\equiv\alpha_s(Q^2)/4\pi$ evolves according to
\begin{equation}
da_s/d\ln Q^2 = -\sum_{k=0}^2 \beta_k a_s^{k+2}
\end{equation}
where $\beta_0=11-2n_f/3\,$, $\,\beta_1=102-38n_f/3\,$ and $\,\beta_2=
2857/2-5033n_f/18$ + $325n_f^2/54$.  Here we utilize the exact numerial 
(iterative) solution for $a_s(Q^2)$ since such an accuracy is mandatory, 
in particular at NNLO, in the low $Q^2$ region relevant for the 
valencelike approach \cite{ref7,ref8}. Since $\beta_k$ is not continuous
for different $n_f$, the continuity of $\alpha_s(Q^2)$ requires to match
$\alpha_s^{(n_f)}$ at $Q=m_h$ $(h=c,b,t$), i.e., $\alpha_s^{(n_f)}(m_h)
=\alpha_s^{(n_f-1)}(m_h)$. These naive matching conditions get corrected
at NNLO \cite{ref23,ref24,ref25} by a marginal term $-(11/72\pi^2)
[\alpha_s^{(n_f-1)}(m_h)]^3$ which will be neglected.  Furthermore we have
chosen $m_c=1.3$ GeV, $m_b=4.2$ GeV and $m_t=175$ GeV, which turned out
to be the optimal choices for our NLO analysis of heavy quark production
\cite{ref8}.

The Mellin $n$-moments of the parton distributions $f(x,Q^2)$,
\begin{equation}
f(n,Q^2) = \int_0^1 dx\, x^{n-1} f(x,Q^2)\,\, ,
\end{equation}
where $f=q,\bar{q},g$, evolve according to
\begin{equation}
\frac{d\vec{q}(n,Q^2)}{d\ln Q^2} = {\hat{P}}(a_s,n)\vec{q}(n,Q^2)
\end{equation}
which refers to the coupled flavor--{\em{singlet}} evolution
equation for $\vec{q}=(\Sigma,g)^T$, $\Sigma= \sum_q(q+\bar{q})$, and
\begin{equation}
\hat{P}(a_s,n)=\sum_{k=0}^2 a_s^{k+1}\,\hat{P}_k(n)=\sum_{k=0}^2 a_s^{k+1}
\left( \begin{array}{cc} 
      P_{qq}^{(k)} & P_{qg}^{(k)}\\
      P_{gq}^{(k)} & P_{gg}^{(k)}\end{array} \right)
\end{equation}
with the well known LO and NLO splitting functions $P_{ij}^{(0)}$ and
$P_{ij}^{(1)}$, respectively, and the NNLO (3-loop) $P_{ij}^{(2)}$ have
been calculated in \cite{ref26}.  Any obvious $Q^2$- and/or $n$-dependence
will be suppressed as far as possible.  The $2\times 2$ matrix evolution
equation (3) can be formally solved recursively \cite{ref27,ref28}
with the result
\begin{equation}
\vec{q}(n,Q^2) = [ \hat{L}+a_s\hat{U}_1\hat{L} -a_{s0}\hat{L}\hat{U}_1
  +a_s^2\hat{U}_2\hat{L} -a_s a_{s0}\hat{U}_1\hat{L}\hat{U}_1
   -a_{s0}^2\hat{L}(\hat{U}_2-\hat{U}_1^2) ]\, \vec{q}(n,Q_0^2)
\end{equation}
where $a_{s0} = \alpha_s(Q_0^2)/4\pi$.  This is the so--called truncated
solution where all redundant ${\cal{O}}(a_s^3)$ terms are disregarded.
The LO evolution operator $\hat{L}=\hat{L}(a_s,a_{s0},n)$, relevant for
the LO solution $\vec{q}_{\rm LO}(n,Q^2) = \hat{L}(a_s,a_{s0},n)\vec{q}
(n,Q_0^2)$, can be written as
\begin{equation}
\hat{L}(a_s,a_{s0},n) 
  \equiv \left (\frac{a_s}{a_{s0}}\right)^{-\hat{R}_0} 
   = \hat{e}_- \left(\frac{a_s}{a_{s0}}\right)^{-\lambda_-} 
    + \hat{e}_+ \left(\frac{a_s}{a_{s0}}\right)^{-\lambda_+}
\end{equation}
where $\hat{R}_0\equiv \hat{P}_0/\beta_0$ and with the projection matrices
$\hat{e}_{\pm}$ being given by
\begin{equation}
\hat{e}_{\pm} = \frac{1}{\lambda_{\pm}-\lambda_{\mp}}\,
    [\hat{R}_0 -\lambda_{\mp}\hat{1}]
\end{equation}
where $\lambda_-(\lambda_+)$ denote the smaller (larger) eigenvalue of
$\hat{R}_0$,
\begin{equation}
\lambda_{\pm} = \frac{1}{2\beta_0}
 [P_{qq}^{(0)}+P_{gg}^{(0)} \pm 
  \sqrt{(P_{qq}^{(0)}-P_{gg}^{(0)})^2 + 4P_{qg}^{(0)} P_{gq}^{(0)}}\,\,]\,\,,
\end{equation}
i.e., $\hat{R}_0 = \lambda_- \hat{e}_-+\lambda_+\hat{e}_+$. Furthermore
\begin{equation}
\hat{U}_{k=1,2} = -\frac{1}{k}
 \left( \hat{e}_- \hat{\tilde{R}}_k \hat{e}_- +\hat{e}_+ 
      \hat{\tilde{R}}_k \hat{e}_+\right)
  + \frac{\hat{e}_+\hat{\tilde{R}}_k\hat{e}_-}{\lambda_--\lambda_+-k}
     +\frac{\hat{e}_- \hat{\tilde{R}}_k \tilde{e}_+}
             {\lambda_+-\lambda_--k}
\end{equation}
with $\hat{\tilde{R}}_{k=1,2} = \hat{R}_k +\sum_{i=1}^{k-1} \hat{R}_{k-i}
\hat{U}_i$ and $\hat{R}_k=\hat{P}_k/\beta_0 -\sum_{i=1}^{k}\beta_i
\hat{R}_{k-i}/\beta_0$.  We have not performed any required matrix
multiplication in (5) analytically, since such a procedure did
not reduce the required computer time of the subsequent numerical analysis.
Rather, we
have performed all required matrix multiplications {\em{entirely}}
{\em{numerically}}, using the $n$-moments of the NNLO splitting 
functions \cite{ref26} $P_{ij}^{(2)}$ appearing in (4) together with the 
standard LO $P_{ij}^{(0)}$ and NLO $P_{ij}^{(1)}$ ones (see, 
e.g.~\cite{ref26}). The Bjorken-$x$ space results $\vec{q}(x,Q^2)$ are 
finally obtained by performing numerically a contour integral around the 
singularities of $\vec{q}(n,Q^2)$ in (5) in the complex $n$-plane in the 
standard way (see, for example, \cite{ref1,ref28,ref29}).

In the flavor--{\em{nonsinglet}} (NS) sector we have a simple
(uncoupled) evolution equation which, in $n$-moment space, reads
\begin{equation}
\frac{dq_{\rm NS}(n,Q^2)}{d\ln Q^2}=P_{\rm NS}(a_s,n)\, q_{\rm NS}(n,Q^2)
\end{equation}
where, similarly to (4), 
$P_{\rm NS}(a_s,n)=\sum_{k=0}^2 a_s^{k+1}P_{\rm NS}^{(k)}(n)$
which refers to the NS splitting functions $P_{\rm NS}^{\pm}$ and
$P_{\rm NS}^v$ (see, for example, \cite{ref30,ref31}).  These splitting
functions govern the evolution of the usual NS combinations of parton
distributions with $q_{\rm NS}$ referring to $q_{\rm NS,\, 3}^{\pm}=
u^{\pm}-d^{\pm}$, $q_{\rm NS,\, 8}^{\pm}=u^{\pm}+d^{\pm}-2s^{\pm}$,
etc., where $q^{\pm}=q\pm\bar{q}$, and $q_{\rm NS}^v=\sum_q(q-\bar{q})$.
The NNLO splitting functions $P_{\rm NS}^{(2)}$ have been given in
\cite{ref31} where the well known LO $P_{\rm NS}^{(0)}$ and NLO 
$P_{\rm NS}^{(1)}$ ones can be found as well.  Since no matrices are
involved in the NS evolution equation (10), its solution can be easily
inferred from the singlet solution (5) where now we have $U_1=-R_1$ and
$2U_2 = -R_2-R_1U_1$ and thus
\begin{equation}
q_{\rm NS}(n,Q^2)=\Big[ 1-(a_s-a_{s0})R_1 
    -\frac{1}{2} (a_s^2-a_{s0}^2)(R_2-R_1^2)
      -a_{s0}(a_s-a_{s0})R_1^2\Big]\, Lq_{\rm NS}(n,Q_0^2)
\end{equation}
with $R_k=P_{\rm NS}^{(k)}/\beta_0-\sum_{i=1}^k \beta_i R_{k-i}/\beta_0$
and $L = L(a_s, a_{s0}, n)=(a_s/a_{s0})^{-R_0}$.  
Again, $q_{\rm NS}(x,Q^2)$ is obtained from a numerical Mellin--inversion
of $q_{\rm NS}(n,Q^2)$.

We have tested our singlet and nonsinglet evolution codes using the 
PEGASUS program \cite{ref32} for generating the  `truncated' solutions
together with the commonly used toy input of the Les Houches and HERA--LHC
Workshops \cite{ref33,ref34}.  For $10^{-7} < x < 0.9$ we achieved an
agreement of up to four decimal places in most cases which is similar to
the required high--accuracy benchmarks advocated in \cite{ref33,ref34}.

As already mentioned at the beginning of this Section we employ for our
analysis the FFNS and fix the number of active light flavors $n_f=3$ in
{\em{all}} splitting functions $P_{ij}^{(k)}$ and in the
corresponding Wilson coefficients to be discussed below.  In this 
factorization scheme only the light quarks ($u,d,s$) are genuine, i.e., 
massless partons within the nucleon, whereas the heavy ones ($c,b,t$) 
are not.  This scheme is fully predictive in the heavy quark sector where
the heavy quark flavors are produced entirely perturbatively from the 
initial light $u,d,s$ quarks and gluons with the full heavy quark mass 
$m_{c,b,t}$ dependence taken into account in the production cross sections 
-- as required experimentally \cite{ref35,ref36,ref37,ref38}, in 
particular in the threshold region.  However, even for very large 
values of $Q^2$, $Q^2\gg m_{c,b}^2$, these FFNS predictions up to NLO
are in remarkable agreement \cite{ref7,ref8} with DIS data and, 
moreover, are perturbatively stable despite the common belief that 
`non--collinear' logarithms $\ln(Q^2/m_h^2)$ have to be resummed for
$h=c,b,t$.

This somewhat questionable resummation of heavy quark mass effects 
using massless evolution equations, starting at the unphysical 
 `thresholds' $Q^2=m_h^2$, is persued in the so--called 
zero--mass  `variable flavor number scheme' (VFNS) where also the
heavy quarks are taken to be massless partons within the nucleon 
with their distributions being generated, e.g.\ up to NLO,
from the boundary conditions
$h(x,m_h^2)=\bar{h}(x,m_h^2)=0$.  Hence this factorization scheme is
characterized by increasing the number $n_f$ of massless partons by
one unit at $Q^2=m_h^2$ starting from $n_f=3$ at $Q^2=m_c^2$, i.e., 
$c(x,m_c^2)=\bar{c}(x,m_c^2)=0$.
The matching conditions are fixed by general continuity relations 
\cite{ref39,ref40} at the respective `thresholds' $Q^2=m_h^2$.
Thus the `heavy' $n_f>3$ quark distributions are perturbatively
uniquely generated from the $n_f-1$ ones via the massless 
renormalization group $Q^2$--evolutions (see, e.g.~[10,15]; a
comparative qualitative and quantitative discussion of the zero--mass
VFNS and the FFNS has been recently presented in \cite{ref41}).
Sometimes one uses an improvement on this, now known as the 
general--mass VFNS 
\cite{ref11,ref12,ref13,ref40,ref42,ref43,ref44,ref45,ref46},
where mass--dependent corrections are maintained in the hard cross
sections.  This latter factorization scheme interpolates between
the zero--mass VFNS and the (experimentally required) FFNS used
for our analysis.

In order to avoid any further dependence on model assumptions, we
choose to work with experimentally directly measurable quantities,
as has been done in \cite{ref8}, like the `reduced' DIS one--photon
exchange cross section $\sigma_r=F_2-(y^2/Y_+)F_L$ together with
the full neutral current (NC) cross sections \cite{ref47}
\begin{equation}
\sigma_{r,\rm NC}^{e^{\pm}p}(x,Q^2) = 
 \left( \frac{2\pi\alpha^2 Y_+}{xQ^4}\right)^{-1}\,
   \frac{d^2\sigma_{\rm NC}^{e^{\pm}p}}{dx\,dQ^2} 
     = F_2^{\rm NC} -\frac{y^2}{Y_+}\, F_L^{\rm NC} \mp
        \frac{Y_-}{Y_+}\, xF_3^{\rm NC}
\end{equation}
where $\alpha= 1/137.036$, $Y_{\pm}=1\pm(1-y)^2$ and
\begin{eqnarray}
F_{2,L}^{\rm NC} & =  & F_{2,L}-v_e\kappa F_{2,L}^{\gamma Z}
    + (v_e^2 + a_e^2)\kappa^2 F_{2,L}^Z\nonumber
\\
F_3^{\rm NC} & = & -a_e\kappa F_3^{\gamma Z} 
    + 2 v_e a_e\kappa^2 F_3^Z\,\, ,
\end{eqnarray}
with $v_e =-\frac{1}{2} +2\sin^2 \theta_W$, $a_e=-\frac{1}{2}$ and
$\kappa^{-1} = 4\sin^2\theta_W\cos^2\theta_W(Q^2+M_Z^2)/Q^2$, using
$\sin^2\theta_W = 0.2312$ and $M_Z = 91.1876$ GeV.  As in our previous
NLO analysis \cite{ref8} it turned out, however, that fitting just to
the usual (one--photon exchange) $F_2(x,Q^2)$ gives rather similar
results.  Defining ${\cal{F}}_{2,L}\equiv F_{2,L}/x$, the $n$-moments
in (2) of these structure functions ${\cal{F}}_{2,L}(x,Q^2)$ and
$F_3(x,Q^2)$ can, for $n_f=3$ light flavors, be written as 
\begin{eqnarray}
{\cal{F}}_{j=2,L}^{\rm NC} & = & C_{j,\rm NS}
 (a_3^+ q_{\rm NS,3}^+ + a_8^+ q_{\rm NS,8}^+) 
   +a^+(C_{j,q}\Sigma + C_{j,g}g)\nonumber
\\
F_3^{\rm NC} & = & C_{3,\rm NS} 
  (a^- q_{\rm NS}^v +a_3^- q_{\rm NS,3}^- + a_8^- q_{\rm NS,8}^-)
\end{eqnarray}
where 
$a_3^{\pm}=\frac{1}{2}(a_u^{\pm}-a_d^{\pm}$), $a^{\pm}_8=\frac{1}{6}
(a_u^{\pm}+a_d^{\pm}-2a_s^{\pm}$) and $a^{\pm}=\frac{1}{3}\sum_{q=u,d,s} 
a^{\pm}_q$ with
\begin{eqnarray}
a_q^+ & = & e_q^2 -2e_q v_e v_q\kappa + (v_e^2 + a_e^2)\kappa^2\nonumber
\\
a_q^- & = & -2e_q a_q\kappa + 4v_e a_e v_q a_q\kappa^2
\end{eqnarray}
where $v_q=\pm \frac{1}{2}-2e_q\sin^2\theta_W$ and $a_q=\pm\frac{1}{2}$
with $\pm$ according to whether q is a $u$- or $d$-type quark. The
Wilson coefficients are generically expanded as \cite{ref30,ref48}
$C_{2,3}(a_s,n) = \sum_{k=0}^2 a_s^k c_{2,3}^{(k)}(n)$ where at LO
$c_{2,\rm NS}^{(0)} = c_{2,q}^{(0)} = c_{3,\rm NS}^{(0)}=1$ and
$c_{2,g}^{(0)}=0$.  The appropriate NLO (1-loop) coefficients 
$c_{2,3}^{(1)}$ can be found, for example, in \cite{ref49}. The NNLO
(2-loop) coefficients $c_{2,3}^{(2)}$ have been originally calculated
in \cite{ref50,ref51} and, for definiteness, we take 
$c_{2,\rm NS}^{(2)}\equiv c_{2,\rm NS}^{(2)+},\, c_{2,q}^{(2)}$ and
$c_{2,g}^{(2)}$ from \cite{ref52}, and $c_{3,\rm NS}^{(2)}\equiv
c_{3,\rm NS}^{(2)-}$ from \cite{ref51}.  Since the longitudinal
structure function $F_L=F_2-2x F_1$ vanishes at LO, it has become
common \cite{ref50} to consider the first nonvanishing
${\cal{O}}(\alpha_s)$ contribution to $F_L$ as the LO one, i.e.,
the perturbative expansion up to NNLO now reads $C_L(a_s,n) =
\sum_{k=1}^3 a_s^k c_L^{(k)}(n)$.  The relevant NNLO (3-loop) 
coefficients $c_L^{(3)}$ have been calculated in \cite{ref52,ref53},
where also the well known LO $c_L^{(1)}$ and NLO $c_L^{(2)}$ 
coefficients can be found.  Although we perform all calculations in
Mellin $n$-moment space, it should be nevertheless mentioned that
in Bjorken-$x$ space the simple products in (14) turn into the 
standard convolutions of the Wilson coefficients with the parton
distributions.

In the medium to large $x$-region the relevant kinematic nucleon target
mass (TM) corrections are also taken into account for the dominant
`light' $F_2$ structure function in (14) (with  `light' referring to
the common $u,d,s$ (anti-)quarks and gluon initiated contributions)
according to \cite{ref54}
\begin{eqnarray}
F_{2,\rm TM}(n,Q^2) & \equiv &
  \int_0^1 dx\, x^{n-2} F_{2,\rm TM}(x,Q^2)\nonumber
\\
& = & \sum_{\ell =0}^2 \left(\frac{m_N^2}{Q^2}\right)^{\ell}
   \frac{(n+\ell)!}{\ell !(n-2)!}\,\, 
    \frac{F_2(n+2\ell ,Q^2)}{(n+2\ell )(n+2\ell -1)} 
     + {\cal{O}}\left( \Big( \frac{m_N^2}{Q^2}\Big)^3\right)
\end{eqnarray}
where higher powers than $(m_N^2/Q^2)^2$ are negligible for the 
relevant $x<0.8$ region, as can straightforwardly be shown by 
comparing (16) with the well--known exact expression in Bjorken-$x$
space \cite{ref54}.

So far we have discussed only the contributions of light partons
($u,d,s,g$) to structure functions in (14), 
$F_{i=2,L,3}^{\rm light}$. The total structure functions 
$F_i(x,Q^2)=F_i^{\rm light}+F_i^{\rm heavy}$ require also the 
knowledge of the (subleading) heavy quark contribution 
$F_i^{\rm heavy} = F_i^c +F_i^b$ at fixed-order of perturbation 
theory. (Top quark contributions are negligible.) 
The LO ${\cal{O}}(\alpha_s)$ contributions to $F_{2,L}^h$, due
to the subprocess $\gamma^*g\to h\bar{h}$ with $h=c,b$, have been
summarized in \cite{ref6}, and the NLO ${\cal{O}}(\alpha_s^2)$ 
ones are given in \cite{ref55,ref56}.  The NNLO ${\cal{O}}
(\alpha_s^3)$ 3-loop corrections to $F_L^h$ and first rudimentary
contributions to $F_2^h$ have been calculated recently 
\cite{ref57,ref58,ref59} for $Q^2\gg m_h^2$, but these asymptotic
results are neither applicable for our present investigation nor
relevant for the majority of presently available data at lower
values of $Q^2$.  Our ignorance of the full NNLO ${\cal{O}}
(\alpha_s^3)$ corrections to $F_{2,L}^h$ constitute the major
problem for any NNLO DIS analysis in the FFNS.  
(It should be
mentioned that we have attempted to mimick the NNLO contributions
by naively assuming them to be down by one power of $\alpha_s$
times the NLO terms multiplied by a constant $K$-factor, but the 
fit results were insensitive to such an ad hoc correction.  
However, this approach (guess) appears to be not appropriate 
since playing the same game at NLO, i.e., $\alpha_s$ times LO
times a $K$-factor can not reproduce the correct NLO results in
the relevant kinematic region of $x$ and $Q^2$.)\footnote{A further
(inconsistent) `check' of the relevance of the unknown massive
NNLO coefficient functions can be made by comparing the predicted
charm and bottom structure functions using our new NNLO (anti)quark
and gluon distributions, as presented in Sec.~3, and the currently
known massive NLO coefficient functions with the fully consistent
NLO predictions (e.g.\ \cite{ref8}) based on NLO parton distributions.
These predictions turn out to be indistinguishable, except at very
small values of $x$, $x$ \raisebox{-0.1cm}{$\stackrel{<}{\sim}$}
$10^{-4}$, where the  `NNLO results' are about 10-15\% smaller than
the NLO ones as shown for example in Figs.~8 and 9 of \cite{ref8}.
This is still fully consistent with the $c\bar{c}$ and $b\bar{b}$
HERA DIS data at very small $x$ \cite{ref35,ref36,ref37,ref38}.}
Therefore, the heavy flavor contributions $F_{2,L}^h$ are taken as
given by fixed order NLO perturbation theory \cite{ref55,ref56}
as in our previous more restricted NNLO analyses \cite{ref60,ref61}.
This is also common in the literature \cite{ref15,ref16,ref17} and
the error in the resulting parton distributions due to NNLO 
corrections to heavy quark production is expected \cite{ref15}
to be less than their experimental errors.
These contributions are gluon $g(x,\mu_F^2)$ dominated and the
factorization scale, also of the remaining parton distributions,
should preferably chosen \cite{ref62} to be $\mu_F^2=4m_h^2$,
although a much larger choice like $\mu_F^2=4(Q^2+4m_h^2)$ leaves
the NLO results essentially unchanged \cite{ref7,ref8}.  The NNLO
heavy quark contributions to $F_3^{\rm NC}$ are not known either,
but here they vanish in LO and are already negligibly small in
NLO at the relevant large values of $Q^2$ as discussed in \cite{ref8}.

More recently the NNLO corrections to the rapidity distribution 
$d^2\sigma/dM\, dy$ of Drell--Yan (DY) dilepton production of mass $M$
has been calculated as well \cite{ref63,ref64}.  This allows to 
include DY data as well for performing a fully consistent analysis
up to NNLO.  Needless to say that the DY $pp$ and $pd$ dilepton
production data are instrumental in fixing $\bar{d}-\bar{u}$ (or
$\bar{d}/\bar{u}$).  Only the usual high-$p_T$ inclusive jet
production data of hadron-hadron scattering have to be disregarded
where the NNLO corrections have not yet been calculated.  The LO
and NLO corrections to the DY process are well known (for a 
summary, see, e.g., \cite{ref65}) and for our full NNLO analysis
we used the routine developed in \cite{ref66} based on the results
of \cite{ref63,ref64}.

Finally, the evaluation of the uncertainties of our NNLO parton
distributions is performed in the same way as of our recent NLO ones
\cite{ref8} which followed the line of \cite{ref67,ref68,ref69}.
The uncertainties $\Delta a_i=a_i -a_i^0$ of the central free fit
parameters $a_i^0$, corresponding to the minimal $\chi_0^2$, are
constrained by $\Delta\chi^2\leq T^2$ with the tolerance parameter
$T$ chosen to be $T^2 = T_{1\sigma}^2=\sqrt{2N}/(1.65)^2\simeq
(4.5)^2$, i.e., $T$ being slightly smaller than in \cite{ref8} due
to the smaller total number of data points considered, $N=1568$,
because the high-$p_T$ jet data and the DIS data for semi-inclusive 
$c\bar{c}$- and $b\bar{b}$-production cannot consistently be included
in a global NNLO fit for the time being.

\section{Quantitative results and dynamical small-$x$ predictions}

Now we extend our recent dynamical LO and NLO($\overline{\rm MS}$)
analysis \cite{ref8} to NNLO($\overline{\rm MS}$).  The valencelike
input distributions $xf(x,Q_0^2)$ at the input scale $Q_0\equiv
\mu<1$ GeV, referring to the flavor nonsinglet (valence) densities
$u_v$, $d_v$, $\Delta\equiv\bar{d}-\bar{u}$ and to the valencelike
densities $\bar{d}+\bar{u}$, $\bar{s}=s$ and $g$ in the singlet
sector, are generically parametrized as 
\begin{equation}
xf(x,Q_0^2) = N_f x^{a_f}(1-x)^{b_f}(1+A_f\sqrt{x}+B_f x)\,\,,
\end{equation}
subject to the constraints $\int_0^1 u_v dx=2, \int_0^1 d_v dx=1$
and
\begin{equation}
\int_0^1x[u_v+d_v+2(\bar{u}+\bar{d}+\bar{s})+g]dx =1\,\, .
\end{equation}
Since the data sets we are using are insensitive to the specific
choice of the strange quark distributions, we continue to generate
the strange densities entirely radiatively \cite{ref7,ref8} starting
from $\bar{s}(x,Q_0^2)=s(x,Q_0^2)=0$ in the dynamical valencelike
approach where $Q_0<1$ GeV. For comparison we also study the common
standard evolution approach, being based on a non--valencelike input
at $Q_0>1$ GeV, where we choose as usual $\bar{s}(x,Q_0^2)=s(x,Q_0^2)
=[\bar{u}(x,Q_0^2)+\bar{d}(x,Q_0^2)]/4$.  Furthermore, since all our 
fits did not require the additional polynomial in (17) for the gluon 
distribution, we have set $A_g=B_g=0$.  This left us with a total of 
21 independent fit parameters, including $\alpha_s$.  As suggested in
\cite{ref10} and done in our previous NLO analysis \cite{ref8}, we
included in our final error analysis only those parameters that are
actually sensitive to the input data set chosen, i.e.\ those 
parameters that are not close  to `flat' directions in the overall
parameter space.  With current data, and our functional form (17),
13 such parameters, including $\alpha_s$, are included in our final
error analysis.  The remaining highly correlated ill--determined eight 
polynomial parameters $A_f$ and $B_f$, with uncertainties of more than 
50\%, were held fixed.

These free parameters have been fixed using the following data sets:
the HERA ep measurements \cite{ref70,ref71,ref72,ref73,ref74} for
$Q^2\geq 2$ GeV$^2$ for the  `reduced' cross sections $\sigma_r$
and $\sigma_{r,\rm NC}$ in (12); the fixed target $F_2^p$ data of
SLAC \cite{ref75}, BCDMS \cite{ref76}, E665 \cite{ref77} and NMC
\cite{ref78}, subject to the standard cuts $Q^2\geq 4$ GeV$^2$ and
$W^2=Q^2(\frac{1}{x}-1)+m_p^2\geq 10$ GeV$^2$, together with the 
structure function ratios $F_2^n/F_2^p$ of BCDMS \cite{ref79}, E665
\cite{ref80} and NMC \cite{ref81}.  Furthermore the Drell--Yan muon
pair production data of E866/NuSea \cite{ref82} for 
$d^2\sigma^{pN}/dM\, dx_F$ with $N=p,d$ have been used as well as
their asymmetry measurements \cite{ref83} for 
$\sigma^{pd}/\sigma^{pp}$.  The DY data are always given in terms 
of $x_F$-distributions, whereas the NNLO expressions have been given
 in terms of the dilepton rapidity $y$-distributions 
\cite{ref63,ref64,ref66}.  Since experimentally the dilepton 
$p_T$ is small
(below about 1.5 GeV) as compared to the dilepton invariant mass
$M$ \raisebox{-0.1cm}{$\stackrel{>}{\sim}$} 5 GeV, we have checked
that it can be safely neglected and the two distributions can be
related using leading order kinematics, as has been done in 
\cite{ref17}: $d^2\sigma/dM\, dx_F=(x_1+x_2)^{-1}d^2\sigma/dM\, dy$
where $x_{1,2}=(M^2/s)^{1/2}e^{\pm y}$, $x_F = x_1-x_2$ and $x_1+x_2=
(x_F^2+4M^2/s)^{1/2}$. All these data sets correspond to 1568 data
points.

The parameters obtained from our NNLO dynamical fit for the input
distributions at the optimal input scale $Q^2_0\equiv\mu_{\rm NNLO}^2
=0.55$ GeV$^2$ are given in Table~1, and the ones for our standard fit,
corresponding to the choice $Q_0^2=2$ GeV$^2$, in Table 2.  Due to
the fact that the higher the perturbative order the faster $\alpha_s
(Q^2)$ increases as $Q^2$ decreases, a NNLO analysis is expected to 
result in a smaller value for $\alpha_s(M_Z^2)$ than a NLO fit in 
order to compensate for this increase. Our dynamical analysis indeed
yields a smaller value at NNLO in Table 1, $\alpha_s^{\rm NNLO}(M_Z^2)
= 0.1124\pm 0.0020$, as compared to our NLO fit
\cite{ref8} which resulted in $\alpha_s^{\rm NLO}(M_Z^2)=0.1145\pm
0.0018$. Both values lie, however, within a $1\sigma$ uncertainty.
(Similar results were obtained in a previous dynamical fit 
\cite{ref61} which was performed for a restricted set of (mainly
small--$x$) DIS data.)  The same holds for our NNLO  `standard' fit 
result in Table 2, $\alpha_s^{\rm NNLO}(M_Z^2) = 0.1158\pm 0.0035$, 
to be compared with the NLO result \cite{ref8} of 
$0.1178\pm 0.0021$.  Our standard NNLO fit result for $\alpha_s(M_Z^2)$ 
in Table 2 agrees, within errors, with the results of \cite{ref15,ref16}
and \cite{ref17}, $0.1143\pm 0.0014$ and $0.1128\pm 0.0015$, 
respectively,
and is compatible with the one obtained from a  `standard' fit 
\cite{ref60} to a restricted set of small--$x$ DIS data. Data for
high--$p_T$ jet production in hadron--hadron scattering should not
be included in a consistent NNLO analysis, since such processes are
theoretically known only up to NLO.  Including them nevertheless
in a standard NNLO analysis requires \cite{ref14,ref84} generally
larger values of $\alpha_s(M_Z^2)$.  Previous standard NNLO fits 
($Q_0>1$ GeV) considering only the flavor non--singlet (NS) valence
sector of structure functions \cite{ref85,ref86,ref87} resulted in
somewhat smaller values of $\alpha_s(M_Z^2)$ than in Table 2 but 
remain within a $1\sigma -2\sigma$ uncertainty.  A similar NS valence
analysis \cite{ref88} as well as a full analysis \cite{ref89} being
based, however, on incomplete calculations of the moments of 3-loop
anomalous dimensions (splitting functions) yielded slightly larger
values of $\alpha_s$ at NNLO, $\alpha_s(M_Z^2)\simeq 0.117$, with
estimated errors large enough so as to comply with our result in 
Table 2.  For a more detailed and comparative recent discussion of
NLO and NNLO results the interested reader is referred to \cite{ref90}.
In general, the NNLO fits result in a better (smaller) $\chi^2$ 
than the NLO ones \cite{ref8}, typically $\chi_{\rm NNLO}^2\simeq
0.9 \chi_{\rm NLO}^2$.

It should be noticed that our $\alpha_s$--uncertainty in Table 2
is about twice as large as the one obtained in a comparable standard
NNLO analysis \cite{ref17} where the high--$p_T$ jet data have been
disregarded for consistency reasons as well.  Without these data the
gluon distribution is little constrained in the medium to large
$x$--region where it plays an important role for the $Q^2$-evolution
at small values of $x$ due to the convolution with the dominant
$P_{gg}^{(k)}$. This $\alpha_s$--uncertainty remains sizeable 
irrespective of the choice of the input scale $Q_0>1$ GeV. Only
within a Bayesian treatment of systematic errors, by taking into
account point--to--point correlations \cite{ref91,ref92,ref93}, the 
uncertainty of $\alpha_s(M_Z^2)$ turns out to be about two times
smaller \cite{ref15,ref16,ref17,ref91,ref92,ref93}. On the other
hand the $\alpha_s$--uncertainty of our dynamical fit in Table 1
is also about half as large as the  `standard' one in Table 2. 
Apart from the larger evolution distance, this is due to the strongly 
constrained valencelike input gluon distribution $xg(x,Q_0^2)=
N_g x^{a_g}(1-x)^{b_g}$ at $Q_0^2=0.55$ GeV$^2$ in the small--$x$
region where $a_g\simeq 1$, according to Table 1; consequently the
energy--momentum sum rule (18) sufficiently constrains $xg(x,Q_0^2)$
in the medium to large $x$--region as we shall see below.

Our dynamical NNLO valence and valencelike (sea and gluon) input
distributions at $Q_0^2=\mu_{\rm NNLO}^2=0.55$ GeV$^2$ are shown
in Fig.~1, according to the parameters in Table 1, together with 
their $1\sigma$ uncertainties.  A comparison
with our previous NLO results \cite{ref8} (dashed curves) shows that
$u_v$ and $d_v$ are now somewhat enhanced around $x=0.1$ to 0.2 and
that a strong and clear valencelike small--$x$ behavior of the NNLO
gluon input is now required, $a_g=0.994\pm 0.379$, as compared to
$a_g\simeq 0.5168 \pm$ 0.4017 at NLO \cite{ref8}.  Furthermore there 
is also
a strong enhancement of the NNLO gluon over the NLO one around
$x=0.1$ and a sizeable depletion at larger values of $x$.  The 
valence distributions of our standard analyses $(Q_0^2=2$ GeV$^2$)
are compared with the standard NNLO ones of Alekhin, Melnikov and
Petriello \cite{ref17} (AMP06, $Q_0^2=9$ GeV$^2$) and the standard
pure NS analysis of Bl\"umlein, B\"ottcher and Guffanti \cite{ref87}
(BBG06, $Q_0^2=4$ GeV$^2$) in Fig.~2 at $Q^2=4$ GeV$^2$. In the 
relevant valence $x$--region, 
$x$ \raisebox{-0.1cm}{$\stackrel{>}{\sim}$} 0.1, 
we confirm the NNLO BBG06 results, in particular the enhancement
of $xd_v$ with respect to the NNLO result of AMP06. In any case we,
as well as BBG06, observe a significant enhancement of the NNLO
$xu_v$ and $xd_v$ with respect to the NLO results.  The valence
distributions are very robust with respect to the choice of the 
input scale $Q_0^2$ since our dynamical valence distributions at
$Q^2=4$ GeV$^2$ practically coincide with the standard ones shown
in Fig.~2. 

The distinctive valencelike gluon input at low $Q^2<1$ GeV$^2$ in
Fig.~1 implies a far stronger constrained gluon distribution at larger
values of $Q^2$ as compared to a gluon density obtained from a 
 `standard' fit with a conventional non--valencelike input at
$Q^2>1$ GeV$^2$, $Q_0^2=2$ GeV$^2$ say, as can be seen in Fig.~3.
In contrast to the standard NLO results \cite{ref8} (shown by the
dotted curves in Fig.~3) the standard NNLO gluon input at 
$Q_0^2=Q^2=2$ GeV$^2$ is very weakly constrained at small $x$ 
($a_g=0.0637 \pm 0.1333$, cf.~Table 2) and therefore (18) cannot
sufficiently constrain it at larger values of $x$ since, moreover,
high--$p_T$ jet data have not been taken into account for 
consistency reasons. Notice that this common standard NNLO input
gluon distribution at $Q_0^2=2$ GeV$^2$ is also compatible with a
valencelike small--$x$ behavior ($a_g>0$) --- a tendency already
observed in \cite{ref13} --- and that our dynamical NNLO gluon
distribution in Fig.~3 (solid curves) remains valencelike even at
$Q^2=2$ GeV$^2$ (i.e.~decreases with decreasing $x$).  This is 
mainly caused by the NNLO splitting function $P_{gg}^{(2)}$ in (4)
which is {\em negative} and {\em more} singular in the small--$x$
region \cite{ref26} than the LO and NLO ones: for $n_f=3$,
$xP_{gg}^{(2)}(x)\sim-3147.66\ln\frac{1}{x} +14737.89$ as $x\to 0$,
whereas $xP_{gg}^{(0)}(x)\sim 12$ and $xP_{gg}^{(1)}(x)\sim-81.33$.
The uncertainties generally decrease as $Q^2$ increases due to the
QCD $Q^2$--evolutions \cite{ref67,ref69}, but the ones of the dynamical
predictions in the small--$x$ region remain substantially smaller
than the uncertainties of the common  `standard' results as
exemplified in Fig.~3.  Furthermore, it is a general feature of 
{\em any} NNLO gluon distribution in the small--$x$ region that it
falls {\em below} the NLO one as can be seen in Fig.~3 by comparing
the solid curves with the long--dashed ones, and the dashed--dotted
curves with the dotted ones. In the first dynamical case the NNLO
predictions for $x<10^{-3}$ are several--$\sigma$ below the NLO
ones.  For comparison we also display in Fig.~3 the standard NNLO
results of AMP06 \cite{ref17} based on an input scale $Q_0^2=9$
GeV$^2$.

The dynamical sea distribution $x(\bar{u}+\bar{d})$ derives from
a less pronounced ($a_{\bar{u}+\bar{d}}<a_g$) valencelike input in
Fig.~1 which vanishes very slowly as $x\to 0$ ($a_{\bar{u}+\bar{d}}
= 0.1374 \pm 0.0501$, cf.~Table 1).  This implies that the 
valencelike sea input is similarly increasing with decreasing $x$
down to $x\simeq 0.01$ as the sea input obtained by the common
standard fit where $a_{\bar{u}+\bar{d}}=-0.1098\pm 0.0122$ according
to Table 2.  Therefore, the $1\sigma$ uncertainty bands of our
dynamically predicted sea distributions at larger values of $Q^2$
in Fig.~4 are only marginally smaller than the corresponding ones
of the standard fit.  In contrast to the evolution of the gluon
distribution in Fig.~3, the NNLO sea distributions in Fig.~4 lie
always {\em above} the NLO ones in the small--$x$ region, 
$x$ \raisebox{-0.1cm}{$\stackrel{<}{\sim}$} $10^{-2}$, and at 
not too large values of $Q^2$.  Here all NNLO sea distributions
are rather similar, including the  `standard' one of AMP06
\cite{ref17}. 

A representative comparison of our dynamical and standard NNLO
results with the relevant HERA(H1,ZEUS) data on the proton structure
function $F_2^p(x,Q^2)$ is presented in Figs.~5 and 6.  It should
be reemphasized that due to our valencelike input, the dynamical
small--$x$ results 
($x$ \raisebox{-0.1cm}{$\stackrel{<}{\sim}$} $10^{-2}$) are
{\em predictions} being entirely generated by the QCD $Q^2$--evolutions.
This is in contrast to a common `standard' fit where the gluon and
sea input distributions in (17) do {\em not} vanish as $x\to 0$.
For comparison we also display our dynamical NLO results \cite{ref8}
shown by the dashed curves.  In all cases the data in Figs.~5
and 6 are well described throughout the whole medium-- to small--$x$
region for 
$Q^2$ \raisebox{-0.1cm}{$\stackrel{>}{\sim}$} 2 GeV$^2$ and thus
perturbative QCD is here fully operative.  At $Q^2<2$ GeV$^2$ the
theoretical results fall below the data in the very small--$x$ 
region; despite the fact that the NNLO results in Fig.~5 are closer
to the data than the NLO ones, this is not unexpected for perturbative
leading twist-2 results since nonperturbative (higher twist)
contributions\footnote{Since operators of different twists do 
{\em{not}} mix under renormalization group evolutions, possible
higher--twist effects do not influence the determination of the input
parameters at the low input scale $Q_0^2=0.55$ GeV$^2$ relevant for
the dynamical (valencelike) leading twist--2 distributions, with the
latter being determined from data at $Q^2\geq 2$ GeV$^2$ and $Q^2\geq 4$
GeV$^2$ with $W^2\geq 10$ GeV$^2$ as discussed above.}
to $F_2(x,Q^2)$ will eventually become relevant, even
dominant, for decreasing values of $Q^2$. Although the inclusion
of NNLO corrections imply an improved value of $\chi^2$, typically
$\chi_{\rm NNLO}^2\simeq 0.0�9 \chi_{\rm NLO}^2$ according to 
\cite{ref8} and Tables 1 and 2, present high precision DIS data are
not sufficiently accurate to distinguish between the NLO results 
and the minute NNLO effects of a few percent.  This is illustrated
in Fig.~7 where the experimental (statistical and systematic) errors
are far bigger than the differences between the NLO and NNLO results.
It should, however, be noticed that the NNLO $1\sigma$ uncertainty
band is somewhat narrower (reduced) than the one at NLO.  The results
are similar for our  `standard' fits.  It has already been noticed
that, by analyzing only the flavor non--singlet valence sector of
structure functions, NNLO effects cannot be delineated by present
data in the medium-- to large--$x$ region, and moreover, 
uncertainties of NLO and LO analyses (such as higher twists, different 
factorization schemes and QED contributions to the QCD 
$Q^2$--evolutions) turn out to be comparable in size to the NNLO 
3-loop contributions \cite{ref86}.

As already pointed out, the measurements of Drell--Yan dilepton 
production in $pp$ and $pd$ collisions \cite{ref82,ref83} are 
instrumental in fixing $\Delta\equiv\bar{d}-\bar{u}$ (or $\bar{d}/
\bar{u}$) \cite{ref94}.  In Figs.~8 and 9 we display our
dynamical NNLO and NLO results, together with the $\pm 1\sigma$
uncertainties, for the differential dimuon mass distributions for
various average values of $x_F=x_1-x_2$ for $pp$ and $pd$ collisions,
respectively.  The `standard' fit results differ only marginally.
In the relevant kinematic region where high--statistics data exist,
all three NNLO and NLO results shown agree within 1$\sigma$. In
Fig.~10 we show the result for the ratio $\sigma^{pd}/2\sigma^{pp}$
relevant for the DY asymmetry $A_{\rm DY}=(\sigma^{pp}-\sigma^{pn})/
(\sigma^{pp}+\sigma^{pn})$. Notice that $\sigma^{pN}\equiv d^2
\sigma^{pN}/dM dx_F\propto \sum_{u,d,s}e_q^2[q(x_1)\bar{q}(x_2)+
q(x_2)\bar{q}(x_1)]$ in LO at a scale $Q^2\equiv M^2$, where $x_1$
and $x_2$ refer to the fractional momenta of the quarks in the beam
($p$) and the nucleon target ($N$), respectively.  Experimentally
$x_F>0$ ($x_1>x_2$) and consequently the Drell--Yan cross section
is dominated by the annihilation of a beam quark with a target
antiquark.

\section{The longitudinal structure function $F_L(x,Q^2)$}

As discussed in Sect.~2 we have explicitly used for our analysis the
experimentally directly measured  `reduced' DIS cross sections (12)
which, for not too large values of $Q^2$, are dominated by the 
one--photon exchange cross section $\sigma_r = F_2-(y^2/Y_+)F_L$
where $y=Q^2/xs$.  The importance of using this quantity has been
emphasized in \cite{ref95}: the effect of $F_L$ becomes increasingly
relevant as $x$ decreases at a given $Q^2$, where $y$ increases. 
This is seen in the data as a flattening of the growth of 
$\sigma_r(x,Q^2)$ as $x$ decreases to very small values, at fixed
$Q^2$, leading eventually to a turnover (cf.~Fig.~11).  At lower
values of $Q^2$ in Fig.~11 ist was not possible in \cite{ref95} to
reproduce this turnover at NLO.  This was mainly due to the negative
longitudinal cross section (negative $F_L(x,Q^2$)) encountered in 
\cite{ref13,ref95}.  Since all of our cross sections and subsequently
all structure functions are manifestly positive throughout the whole
kinematic region considered, our dynamical NLO \cite{ref8} and NNLO
results in Fig.~11 are in good agreement with all small--$x$ HERA
measurements [70--74].
The same holds true for our
`standard' NLO \cite{ref8} and NNLO results which, besides having
slightly wider uncertainty bands, are almost indistinguishable from
the dynamical ones shown in Fig.~11.  In Fig.~12 we display our 
NNLO results for $\sigma_r$ at different proton beam energies $E_p$,
relevant for most recent H1 measurements \cite{ref96}, where the
turnover at small $x$ becomes more pronounced at smaller energies
because of the larger values of $y$.  Our dynamical small--$x$
predictions are fully compatible with the (preliminary) H1 data 
presented in \cite{ref96}.

Turning now to $F_L$ itself we note that the $n$--moment equation
(14) for $F_L$ becomes in Bjorken--$x$ space
\begin{equation}
x^{-1}F_L(x,Q^2) = C_{L,NS}\otimes
 \Big(\frac{1}{6}\ q_{NS,3}^+ +\frac{1}{18}\, q_{NS,8}^+\Big)
  +\frac{2}{9} \left( C_{L,q}\otimes\Sigma + C_{L,g}\otimes g\right)
   + x^{-1}F_L^{\rm heavy}(x,Q^2) 
\end{equation}
where $\otimes$ in the light parton sector denotes the common 
convolution, and the weak $Z^0$ contributions in (13) and (15) have
been neglected for $Q^2\ll M_Z^2$ relevant for our present interest.
We have also added the heavy quark (charm, bottom) contribution 
$F_L^{\rm heavy}$ for which we use, as discussed in Sect.~2, the
NLO(2--loop) expressions also in
NNLO due to our ignorance of the ${\cal{O}}(\alpha_s^3)$ NNLO 
heavy quark corrections.  (This inconsistency is here of minor
importance since $F_L^c$ -- and even more so $F_L^b$ -- is a 
genuinely subdominant NLO contribution to the total $F_L$, which 
holds of course also at LO (cf.~Figs.~13 and 14)). Following the
notation of Sect.~2, the perturbative expansion up to NNLO of the
coefficient functions in (19) reads $C_{L,i}(\alpha_s,x) =
\sum_{k=1}^3 a_s^k c_{L,i}^{(k)}(x)$.  In LO, $c_{L,NS}^{(1)}=
\frac{16}{3}x$, $c_{L,ps}^{(1)}=0$, $c_{L,g}^{(1)}=24x(1-x)$ and
the flavor--singlet quark coeefficient function is decomposed 
into the non--singlet and `pure singlet' contribution, 
$c_{L,q}^{(k)}=c_{L,NS}^{(k)}+c_{L,ps}^{(k)}$.  Sufficiently
accurate simplified expressions for the NLO \cite{ref97,ref98,ref99}
and NNLO \cite{ref52} coefficient functions $c_{L,i}^{(2)}$ and
$c_{L,i}^{(3)}$, respectively, have been given in \cite{ref53}. 
It has been furthermore noted in \cite{ref53} that especially for
$C_{L,g}$ both NLO and NNLO contributions are rather large over
almost the entire $x$--range.  Most striking, however, is the 
behavior of both singlet coefficient functions $C_{L,q}$ and 
$C_{L,g}$ in (19) at very small values of $x$: the vanishingly 
small LO parts ($xc_{L,i}^{(1)}\sim x^2$) are negligible as 
compared to the negative constant NLO 2--loop terms, which in turn
are completely overwhelmed by the {\em positive} NNLO 3--loop 
singular contribution $xc_{L,i}^{(3)}\sim\ln\frac{1}{x}$. This
latter singular correction might be indicative for a perturbative
instability at NNLO \cite{ref53} but it should be kept in mind
that a small--$x$ information alone is {\em insufficient} for 
reliable estimates of the convolutions occuring in (19) when 
evaluating physical observables.

Our dynamical LO, NLO and NNLO predictions for the total $F_L$ are
displayed in Fig.~13, together with the small subdominant charm
contributions at LO and NLO.  
These predictions become perturbatively stable already at 
$Q^2$ \raisebox{-0.1cm}{$\stackrel{>}{\sim}$} 3 GeV$^2$  where the 
gluon contribution becomes dominant and
where precision measurements could even delineate NNLO effects
in the very small--$x$ region.  
It should be noted that the error bands at smaller values of $Q^2$ 
(where the gluon and quark contributions are still comparable) are
rather small.  This is caused by compensating effects in the error
analysis, although the individual light quark and gluon contributions
have wider error bands.  Within the common `standard' approach the
absolute values of the NNLO and NLO results in Fig.~14 differ by more
at smaller $Q^2$ than the dynamical ones, 
but the differences between the NNLO and NLO results
are here less distinguishable due to the larger $1\sigma$ 
uncertainty bands which partly overlap in the very small--$x$
region.  
It should, however, be noted that it is somewhat deceptive to 
compare the error bands in Figs.~13 and 14: in the  `dynamical'
approach (Fig.~13) the errors are strongly reduced due to the 
evolution from about 0.5 GeV$^2$, in contrast to the much smaller
evolution distances in the  `standard' approach (Fig.~14) where
the evolution starts at the input scale $Q_0^2=Q^2=2$ GeV$^2$.
It should furthermore be noticed that the NLO/NNLO instabilities
implied by the standard fit results obtained in \cite{ref13,ref95}
at $Q^2$ \raisebox{-0.1cm}{$\stackrel{<}{\sim}$} 5 GeV$^2$ are
far more violent than the ones shown in Fig.~14 which is mainly
due to the negative longitudinal cross section (negative $F_L(x,Q^2)$)
encountered in \cite{ref13,ref95}.  The perturbative stability in 
any scenario becomes in general better the larger $Q^2$, typically
beyond $4-5$ GeV$^2$ \cite{ref13,ref53,ref61,ref95}, as evident
from Figs.~13 and 14.  This is due to the fact that the 
$Q^2$-evolutions eventually force any parton distribution to 
become sufficiently steep in $x$.  It should be mentioned that
the sizeable discrepancies between NNLO and NLO predictions at
$Q^2 = 2$ GeV$^2$ and $x\simeq 10^{-5}$ in Figs.~13 and 14 are not
too surprising since $Q^2\simeq 2$ GeV$^2$ represents somehow a 
borderline value for the leading twist--2 contribution to become
dominant at small--$x$ values.  This is further corroborated by 
the observation that the dynamical NNLO and NLO twist--2 fits 
slightly undershoot the HERA data for $F_2$ at 
$Q^2$ \raisebox{-0.1cm}{$\stackrel{<}{\sim}$} 2 GeV$^2$ in the
small--$x$ region (cf.~Fig.~5), which indicates that 
nonperturbative (higher twist) contributions$^2$ to structure 
functions become relevant for 
$Q^2$ \raisebox{-0.1cm}{$\stackrel{<}{\sim}$} 2 GeV$^2$ 
\cite{ref7,ref8}.

For completeness we finally compare in Fig.~15 our NNLO dynamical
and standard (leading twist) predictions for $F_L(x,Q^2)$, together 
with their $\pm 1\sigma$ error bands, with a representative
selection of (partly preliminary) H1 data 
\cite{ref72,ref73,ref100,ref101,ref102} at fixed $W\simeq 276$
GeV.  For comparison we also show in Fig.~15 our NLO results
\cite{ref8} which have $1\sigma$ uncertainty bands similar to
the NNLO ones.  All our NNLO and NLO results for $F_L$, being
gluon dominated in the small--$x$ region, are in full agreement
with present measurements which is in contrast to expectations
\cite{ref12,ref13,ref95} based on negative parton distributions
and structure functions at small values of $x$.  To illustrate
the manifest positive definiteness of our dynamically generated
structure functions at $Q^2\geq \mu^2$ ($\mu_{\rm NNLO}^2=0.55$
GeV$^2$, $\mu_{\rm NLO}^2=0.5$ GeV$^2$), we show $F_L(x,Q^2)$
in Fig.~15 down to small values of $Q^2$ although leading
twist--2 predictions need not necessarily be confronted with
data below, say, 2 GeV$^2$.

\section{Summary and conclusions}

Utilizing recent DIS structure function measurements ($F_{2,3,L}$
and the   `reduced' cross section $\sigma_r$) and hadronic
Drell--Yan dilepton production data, our previous LO and NLO
global fit analyses for the dynamical parton distributions of
the nucleon \cite{ref8} have been extended to NNLO of
perturbative QCD.  The small--$x$ 
($x$ \raisebox{-0.1cm}{$\stackrel{<}{\sim}$} $10^{-2}$) structure
of dynamical parton distributions is generated entirely 
radiatively from {\em valencelike}, manifestly {\em positive},
input distributions at an optimally chosen input scale $Q_0^2<1$
GeV$^2$.  The NNLO predictions are perturbatively stable with
respect to the NLO ones and are in agreement with all present
measurements for 
$Q^2$ \raisebox{-0.1cm}{$\stackrel{>}{\sim}$} 2 GeV$^2$. 
In general, the NNLO corrections imply an improved value of
$\chi^2$, typically $\chi_{\rm NNLO}^2\simeq 0.9$ 
$\chi_{\rm NLO}^2$.
Having augmented our analyses with an appropriate uncertainty
analysis, it turned out that present DIS precision data are still
not sufficiently accurate to distinguish between NLO results and
the minute NNLO effects of a few percent, despite of the fact
that the dynamical NNLO $1\sigma$ uncertainties are somewhat
smaller than the NLO ones. Inclusive high--$p_T$ jet data were
disregarded for consistency reasons, since NNLO corrections have
not yet been calculated.  Nevertheless the {\em valencelike}
input gluon distribution remains sufficiently constrained, via
the energy--momentum sum rule, also at larger values of $x$
(which is in contrast to a common  `standard' fit approach).
It is interesting to note that our dynamical NNLO gluon 
distribution remains valencelike even at $Q^2=2-3$ GeV$^2$ (i.e.\
decreases with decreasing $x$, cf.~Fig.~3) which is mainly caused
by the dominant NNLO gluon--gluon  splitting function 
$P_{gg}^{(2)}$ which is negative and more singular as $x\to 0$
than the LO and NLO ones, $P_{gg}^{(2)}(x)\!\!\!~\sim -\frac{1}{x}
\ln\frac{1}{x}$.  The drawback of any precision NNLO analysis at
present is that the experimentally required heavy quark mass
effects of heavy quark (charm, bottom) contributions can only
be taken into account up to NLO because of our ignorance of the
full NNLO ${\cal{O}}(\alpha_s^3)$ corrections.   

Our dynamical distributions and predictions have also been
compared with conventional (`standard') NNLO ones obtained from
{\em non}--valencelike positive definite input distributions at
some arbitrarily chosen higher input scale $Q_0^2>1$ GeV$^2$. 
For this purpose we have performed a common `standard' fit as
well, assuming $Q_0^2=2$ GeV$^2$ (notice that, contrary to the
dynamical approach, the finite small--$x$ behavior of the input
gluon and sea distributions is here fitted, and {\em not}
dynamically generated by QCD evolutions). As in the dynamical
approach, the NNLO corrections imply here also an improved
$\chi^2$, typically $\chi_{\rm NNLO}^2\simeq 0.9\, 
\chi_{\rm NLO}^2$. The $1\sigma$ uncertainties of these less
constrained   `standard' distributions are, as expected, larger
than those of their dynamical counterparts, in particular in
the small--$x$ region.

Our predictions for the longitudinal structure $F_L(x,Q^2)$ and
results for the  `reduced' DIS cross section $\sigma_r(x,Q^2)$
are in agreement with all HERA data and most recent HERA--H1
measurements, in particular in the small--$x$ region and down
to $Q^2=2$ GeV$^2$.  The dynamical NNLO/NLO predictions for
$F_L(x,Q^2)$ become perturbatively stable already at $Q^2=2-3$
GeV$^2$ where future precision measurements could even delineate
NNLO effects in the very small--$x$ region (the NNLO $1\sigma$
uncertainty bands are here smaller than the NLO ones).  This is
in contrast to the common `standard' approach but NNLO/NLO
instabilities and differences are here less distinguishable due
to the much larger $1\sigma$ error bands.

The strong coupling obtained from our dynamical NNLO analysis
is $\alpha_s^{\rm NNLO}(M_Z^2)=0.1124\pm 0.0020$ to be compared
with $\alpha_s^{\rm NLO}(M_Z^2)=0.1145\pm 0.0018$ at NLO
\cite{ref8}.  The less constrained  `standard' approach at NNLO
resulted in $\alpha_s^{\rm NNLO}(M_Z^2)= 0.1158\pm 0.0035$ to
be compared with $\alpha_s^{\rm NLO}(M_Z^2)= 0.1178\pm 0.0021$
at NLO \cite{ref8}.

For our analysis in the `fixed flavor number scheme' with $n_f=3$
active light ($u,d,s$) flavors, we have developed our own 
(entirely numerical) NNLO $Q^2$--evolution algorithm which has
been tested by reproducing the appropriate Les Houches and 
HERA--LHC high--accuracy benchmarks \cite{ref33,ref34}. 
A FORTRAN code (grid) containing our NNLO dynamical and `standard'
light ($u,d,s;g)$ parton distributions can be obtained on request
or directly from {\tt http://doom.physik.uni-dortmund.de/pdfserver}.
\vspace{1.0cm}

\noindent{\large{\bf Acknowledgements}}\\
We thank very much S.I.~Alekhin for providing us with a routine
for calculating the NNLO corrections to the rapidity distribution
of Drell--Yan dilepton production, as well as for a clarifying 
correspondence, and J.~Bl\"umlein for a helpful correspondence.
We are very indebted to M.~Gl\"uck for many discussions and for
his continued interest. This work has been supported in part by 
the  `Bundesministerium f\"ur Bildung und Forschung', Berlin/Bonn.

\newpage
\begin{table}[t]
\renewcommand{\arraystretch}{1.8} 
\vskip 14pt
\scriptsize
\centering
\begin{center}
\begin{tabular}{|c||c|c|c|c|c|}
\hline
&\multicolumn{5}{|c|}{NNLO ($\overline{\rm MS}$)} \\
\hline
& $u_v$ & $d_v$ & $\bar{d} - \bar{u}$ & $\bar{u} + \bar{d}$ & $g$ \\
\hline
N & 4.4049 & 13.824 & 8.6620 & 1.2316 & 23.034 \\ 

a & 0.7875 & 1.1778 & 1.2963 & 0.1374 & 0.9940 \\ 

b & 3.6857 & 5.6754 & 19.057 & 10.843 & 6.7892 \\ 

A & -1.1483 & -2.2415 & -6.8745  & -4.5634 & - \\ 

B & 4.5921 & 3.5917 & 19.402 & 11.940 & - \\ 

\hline
$\chi^2/{\rm dof}$ & \multicolumn{5}{|c|}{0.986 (0.904)} \\ 
\hline
$\alpha_s(M_Z^2)$ & \multicolumn{5}{|c|}{0.1124 $\pm$ 0.0020} \\ 
\hline
\end{tabular}
\vspace{1em}
\normalsize
\caption{Parameters of our dynamical input distributions as 
parametrized in (17) referring to an input scale of $Q_0^2\equiv
\mu_{\rm NNLO}^2=0.55$ GeV$^2$.  Since the gluon distribution 
turned out to be insensitive to the polynomial terms in (17), we
have set them to zero ($A_g=B_g=0$). The total numbers of degrees
of freedom is ${\rm dof}=1568-21=1547$. The $\chi^2/$dof in 
brackets refers just to the DIS data where ${\rm dof}=1178-21=
1157$. Furthermore $\alpha_s(\mu_{\rm NNLO}^2)/\pi=0.1522$.} 
\end{center}
\end{table}
\begin{table}[ht]
\renewcommand{\arraystretch}{1.8} 
\vskip 14pt
\scriptsize
\centering
\begin{center}
\begin{tabular}{|c||c|c|c|c|c|}
\hline
&\multicolumn{5}{|c|}{NNLO ($\overline{\rm MS}$)} \\
\hline
& $u_v$ & $d_v$ & $\bar{d} - \bar{u}$ & $\bar{u} + \bar{d}$ & $g$ \\
\hline
N & 3.2350 & 13.058 & 8.1558 & 0.4250 & 3.0076 \\ 

a & 0.6710 & 1.0701 & 1.1328 & -0.1098 & 0.0637 \\ 

b & 3.9293 & 6.2177 & 21.043 & 10.341 & 5.4473 \\ 

A & -0.5302 & -2.5830 & -7.6334  & -3.0946 & - \\ 

B & 3.9029 & 3.8965 & 20.054 & 11.613 & - \\ 

\hline
$\chi^2/{\rm dof}$ & \multicolumn{5}{|c|}{0.947 (0.873)} \\ 
\hline
$\alpha_s(M_Z^2)$ & \multicolumn{5}{|c|}{0.1158 $\pm$ 0.0035} \\ 
\hline
\end{tabular}
\vspace{1em}
\normalsize
\caption{As Table 1 but for the input parameters in (17) of the
standard fit at an input scale $Q_0^2=2$ GeV$^2$, where
$\alpha_s(Q_0^2)/\pi = 0.1072$.}
\end{center}
\end{table}
\clearpage
\begin{figure}
\begin{center}
\ifpdf
\includegraphics[width=14.0cm]{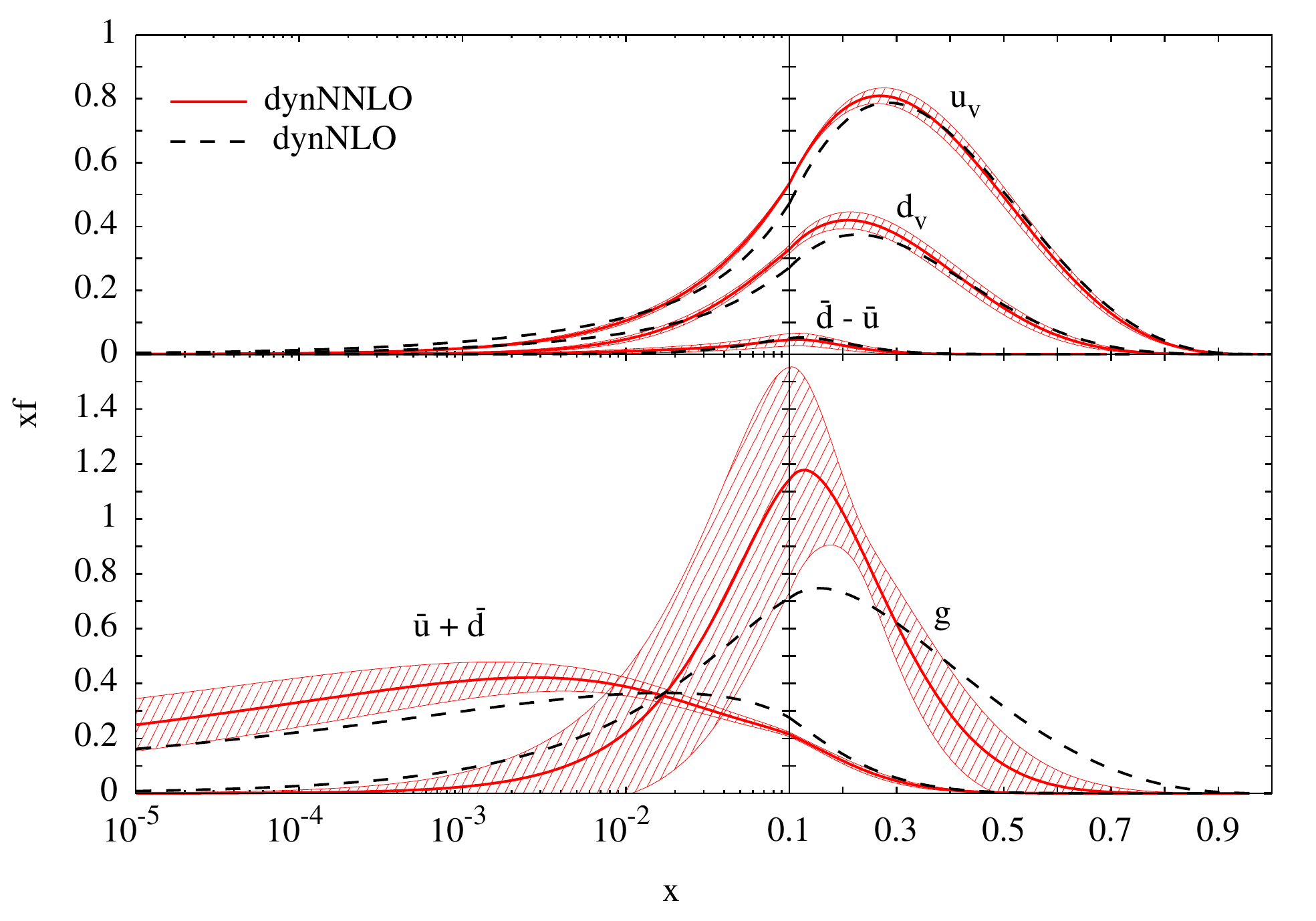}
\fi
\caption{The dynamical NNLO valence ($u_v,d_v,\bar{d}-\bar{u}$) and
valencelike ($g,\bar{u}+\bar{d}$) input distributions together 
with their $\pm 1\sigma$ uncertainties at $Q_0^2\equiv\mu_{\rm NNLO}^2
=0.55$ GeV$^2$.  The central curves follow from (17) with the 
parameters given in Table 1. The strange sea $s=\bar{s}$ vanishes
at the input scale.  Our dynamical NLO input \cite{ref8} at
$Q_0^2\equiv\mu_{\rm NLO}^2=0.5$ GeV$^2$ is also shown by the 
dashed curves for comparison.  The $1\sigma$ uncertainties at
NLO \cite{ref8} are comparable to the ones shown at NNLO, except
for the NLO gluon at 
$x \gtrsim 0.3$
which is stronger constrained due to the light high--$p_T$ jet
data \cite{ref8}.}
\end{center}
\end{figure}
\clearpage
\begin{figure}
\begin{center}
\ifpdf
\includegraphics[width=14.0cm]{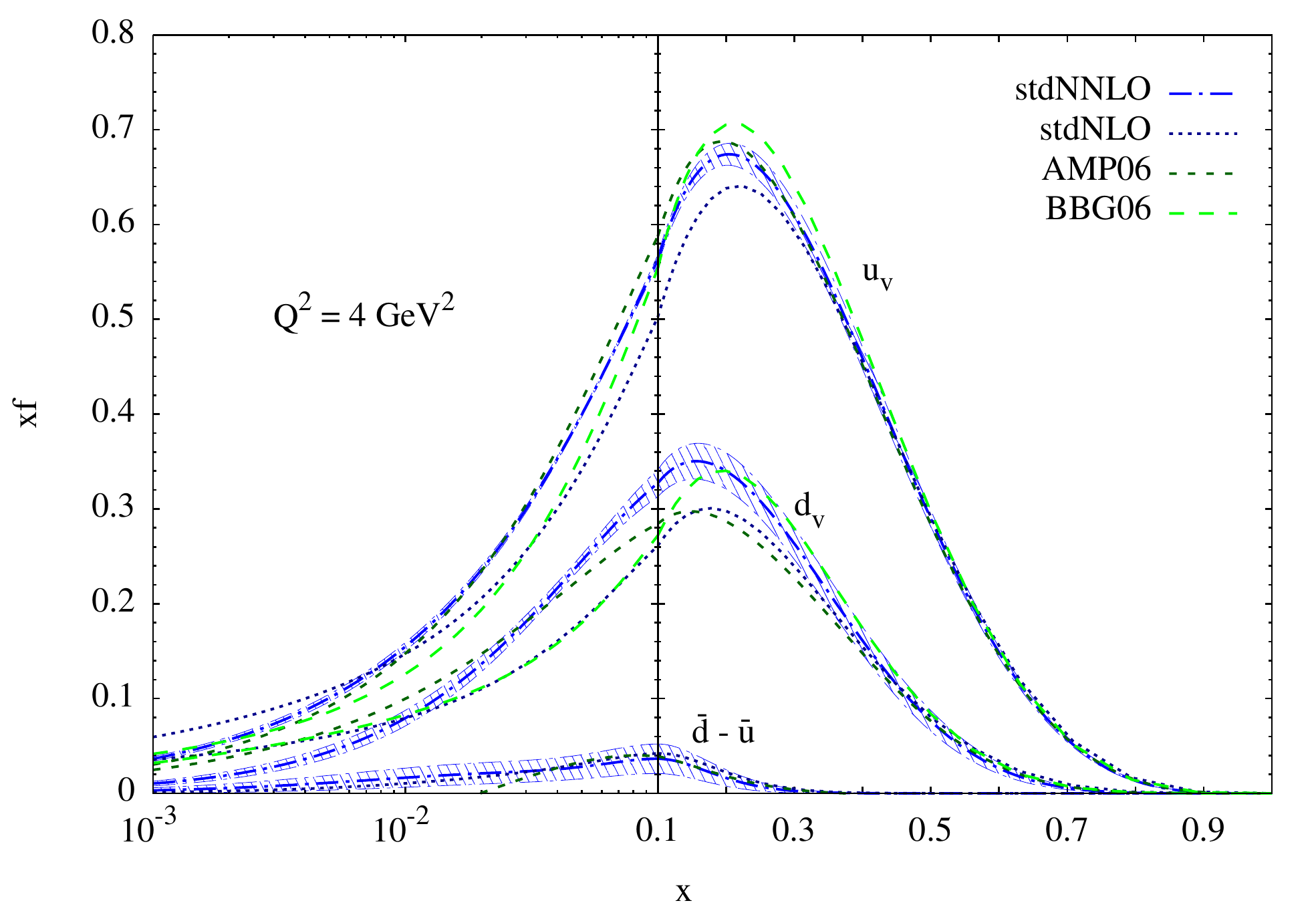}
\fi
\caption{Our standard NNLO valence dsitributions together with their
$\pm 1\sigma$ uncertainties at $Q^2=4$ GeV$^2$, according to the input
parameters in Table 2 at $Q_0^2=2$ GeV$^2$ for the central curves.
Our standard NLO results \cite{ref8} are shown by the dotted curves.
For comparison the standard NNLO results of AMP06 \cite{ref17} and
BBG06 \cite{ref87} are shown as well.  Our dynamical valence 
distributions at $Q^2=4$ GeV$^2$ practically coincide with the 
standard ones shown.} 
\end{center}
\end{figure}
\clearpage
\begin{figure}
\begin{center}
\ifpdf
\includegraphics[width=14.0cm]{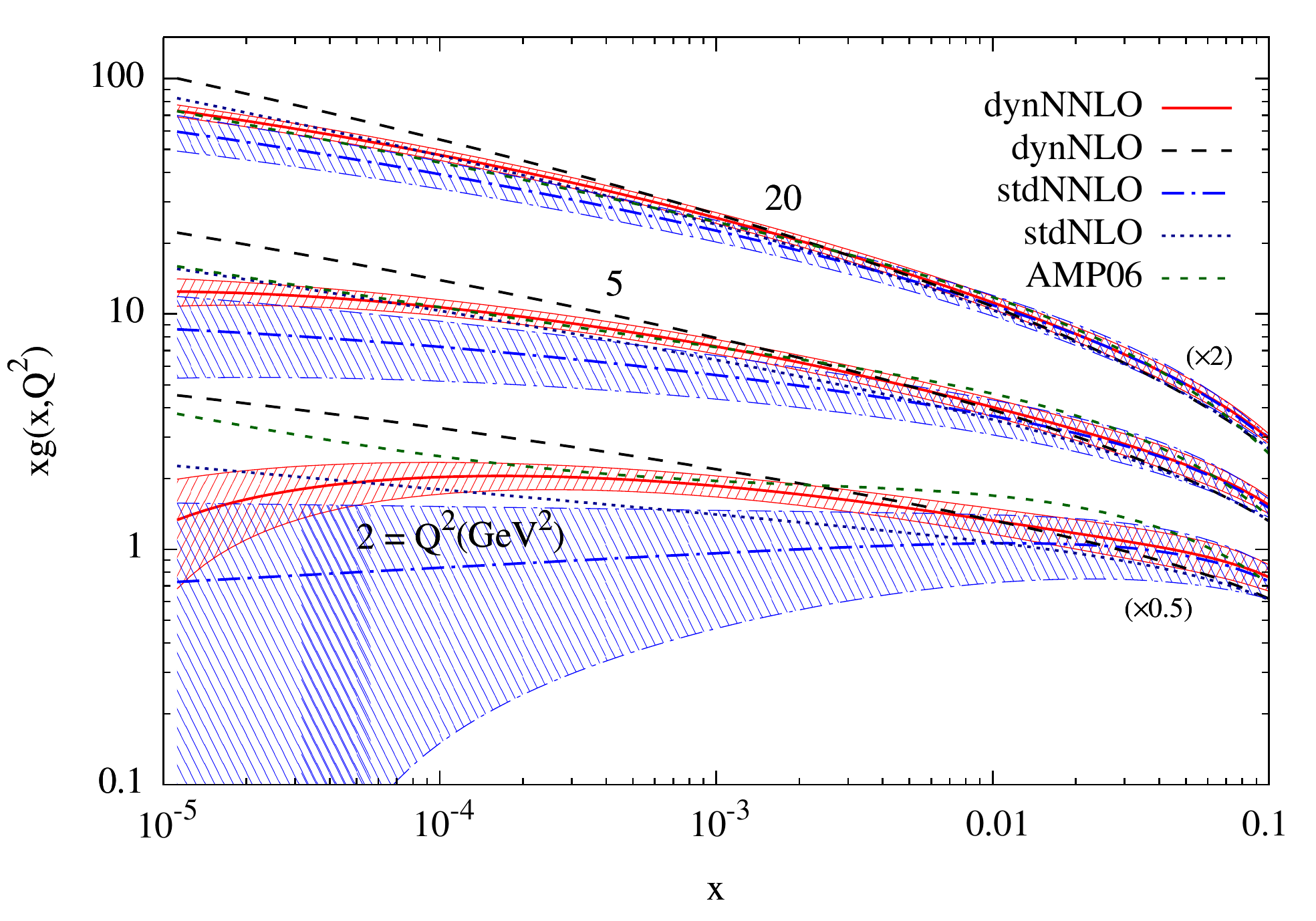}
\fi
\caption{Comparing the $\pm 1\sigma$ error bands of our dynamical (dyn)
and standard (std) NNLO gluon distributions at small $x$ for various
fixed values of $Q^2$.  Note that $Q^2=2$ GeV$^2$ is the input scale
of the standard fit.  The central NLO results are taken from \cite{ref8}
with uncertainties comparable to the ones shown for NNLO for $Q^2$
above 2 GeV$^2$.  For comparison the  `standard' NNLO results of 
AMP06 \cite{ref17} are shown as well.  The results at $Q^2=2$ and 20
GeV$^2$ have been multiplied by 0.5 and 2, respectively, as indicated
in the figure.} 
\end{center}
\end{figure}
\clearpage
\begin{figure}
\begin{center}
\ifpdf
\includegraphics[width=14.0cm]{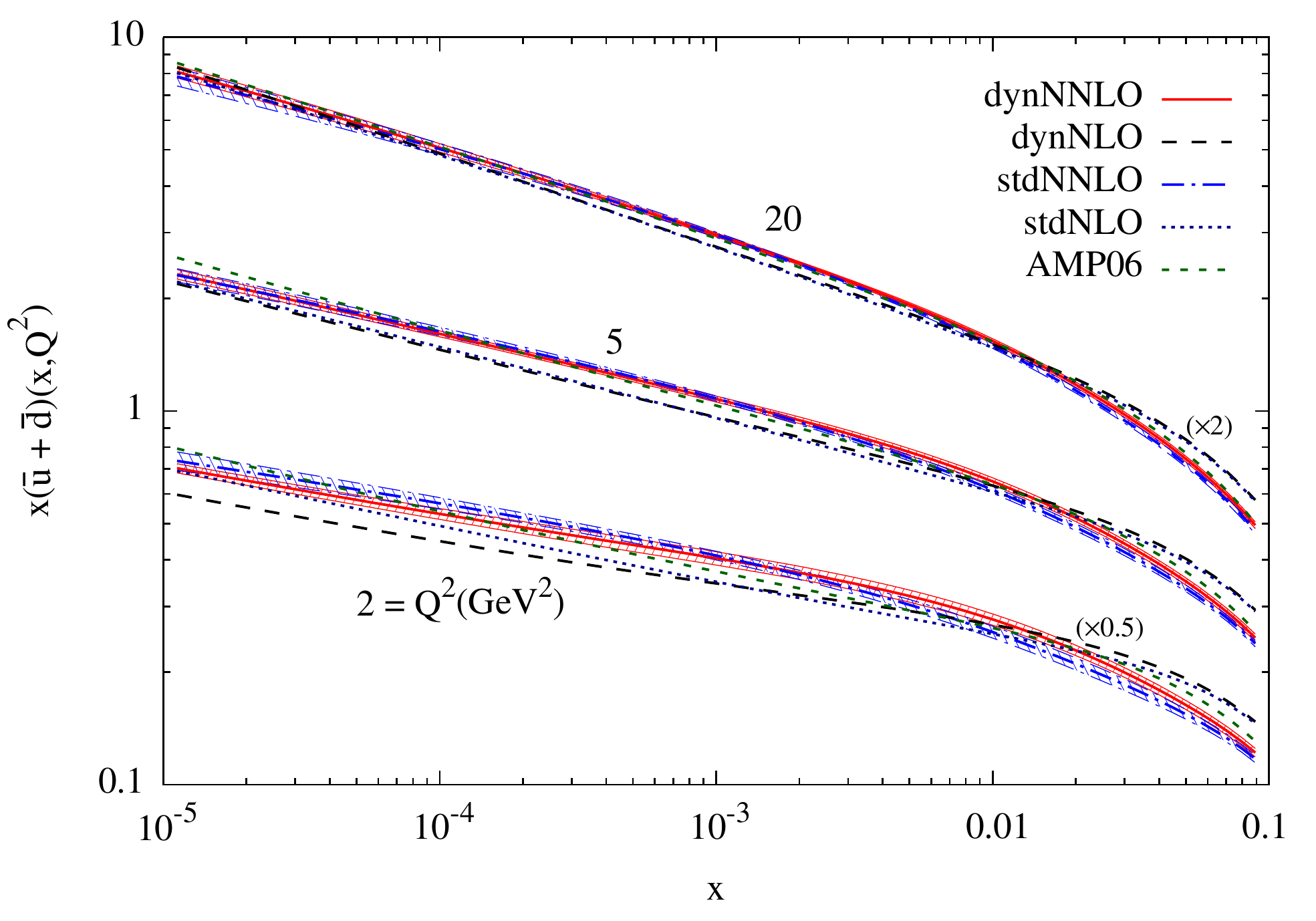}
\fi
\caption{As in Fig.~3 but for the sea quark distribution 
$x(\bar{u}+\bar{d}$).}
\end{center}
\end{figure}
\clearpage
\begin{figure}
\begin{center}
\ifpdf
\includegraphics[width=16.0cm]{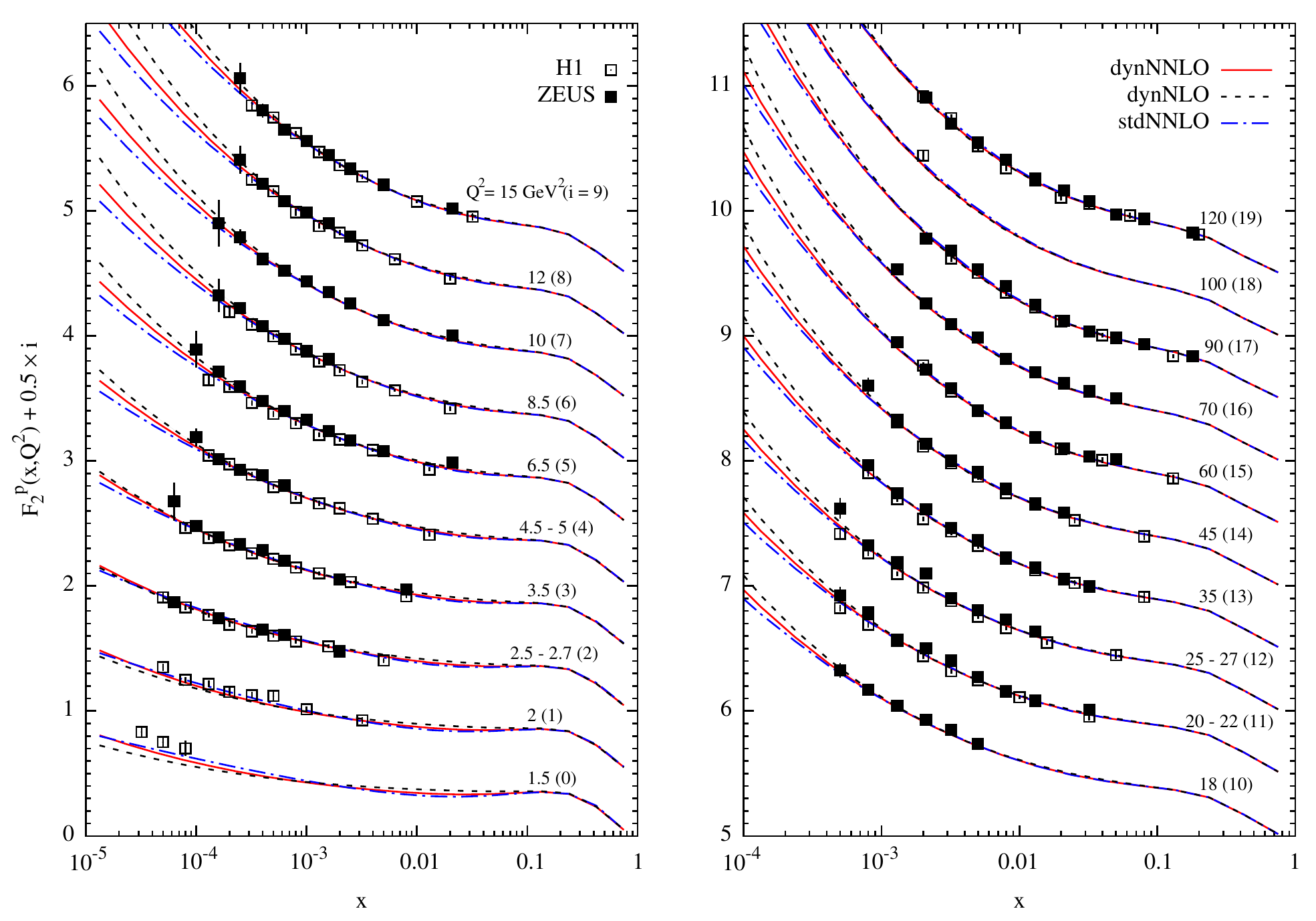}
\fi
\caption{Comparison of our dynamical (dyn) and standard (std) NNLO 
small--$x$ results for $F_2^p(x,Q^2)$ with HERA data for $Q^2\geq 1.5$
GeV$^2$ [70--74]. The dynamical NLO results are taken from \cite{ref8}.
To ease the graphical presentation we have plotted $F_2^p(x,Q^2)+
0.5i(Q^2)$ with $i(Q^2)$ indicated in parentheses in the figure for
each fixed value of $Q^2$.}
\end{center}
\end{figure}
\clearpage
\begin{figure}
\begin{center}
\ifpdf
\includegraphics[width=16.0cm]{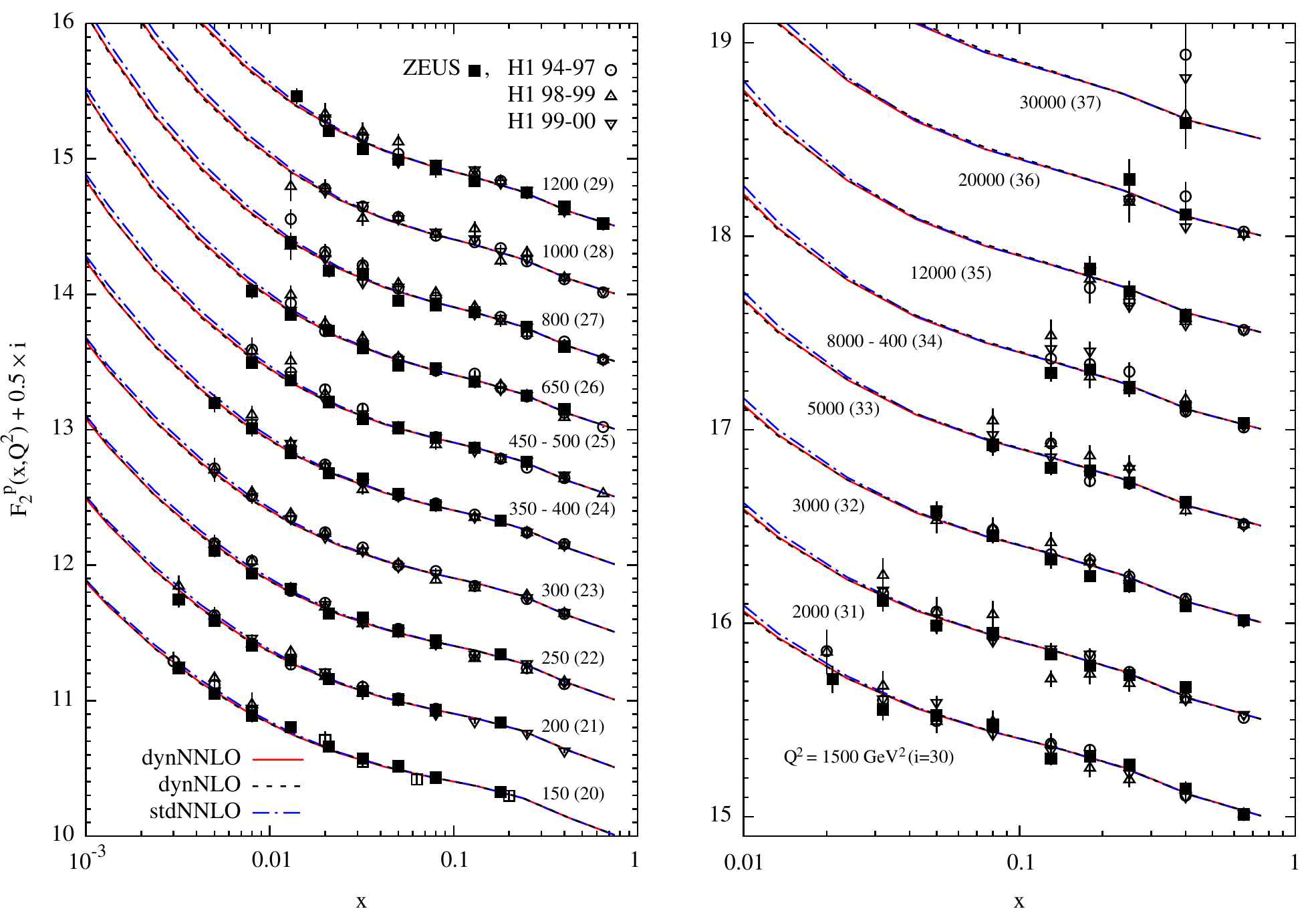}
\fi
\caption{As in Fig.~5 but for large values of $Q^2$ and larger $x$.}
\end{center}
\end{figure}
\clearpage
\begin{figure}
\begin{center}
\ifpdf
\includegraphics[width=14.0cm]{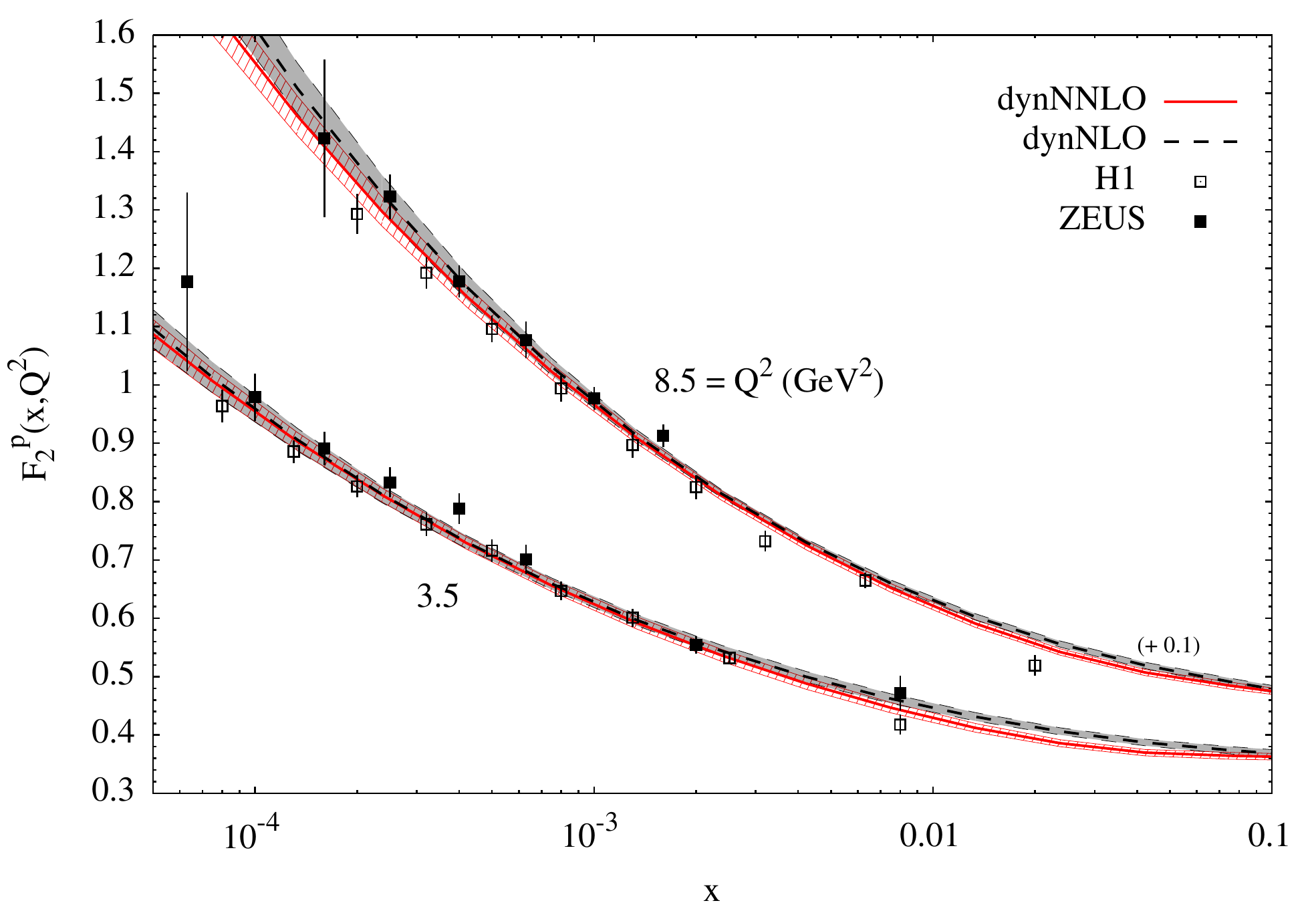}
\fi
\caption{Typical $\pm 1\sigma$ uncertainty bands of our dynamical NNLO
and NLO results in Fig.~5 for two representative values of $Q^2$. The
`standard' NNLO and NLO results are very similar.  To ease the 
visibility we have added 0.1 to the results for $Q^2=8.5$ GeV$^2$ as
indicated.}
\end{center}
\end{figure}
\clearpage
\begin{figure}
\begin{center}
\ifpdf
\includegraphics[width=16.0cm]{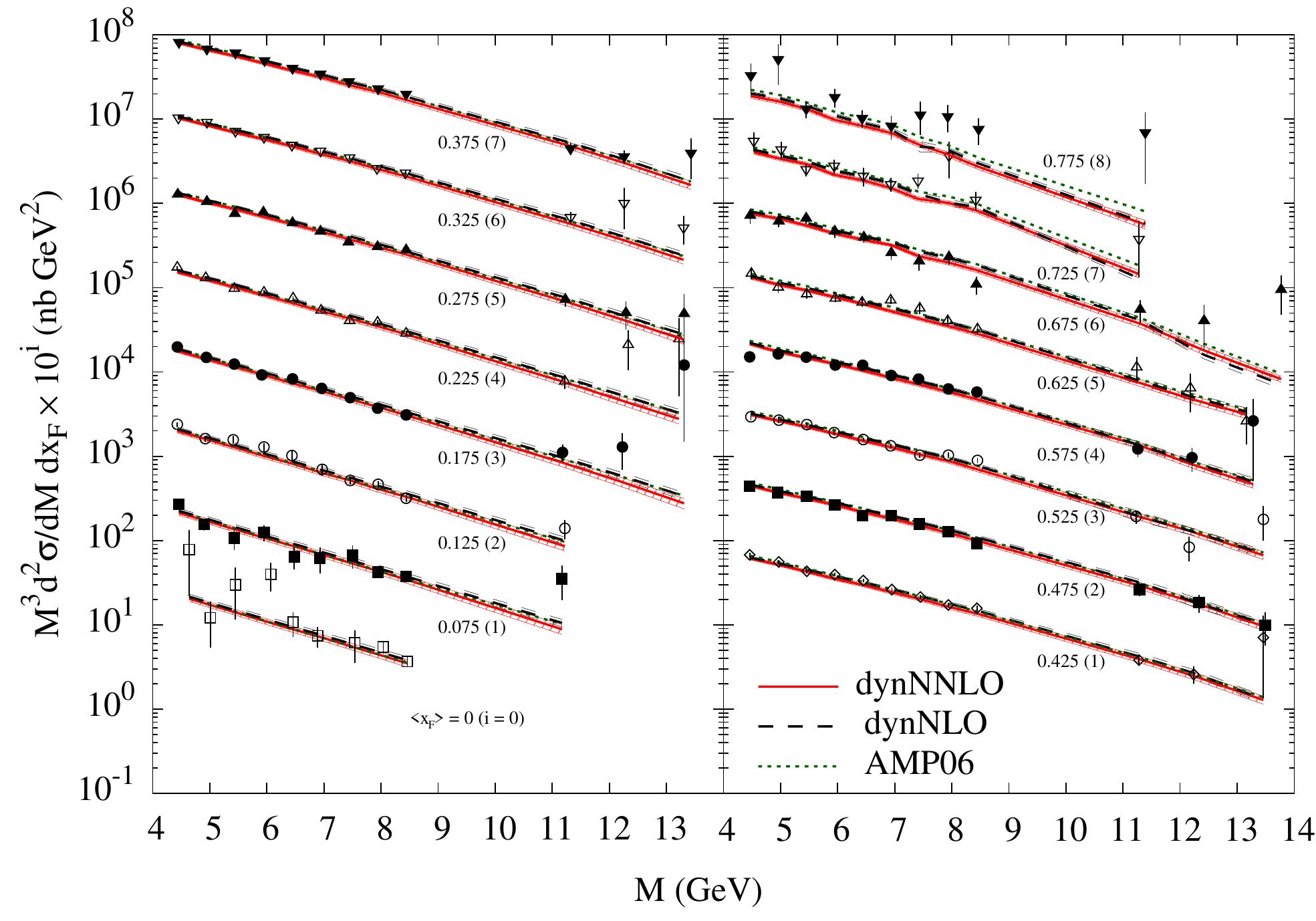}
\fi
\caption{Our dynamical NNLO and NLO \cite{ref8} results, together with
their $\pm 1\sigma$ uncertainties, for Drell--Yan dilepton production 
in $pp$ collisions for various selected average values of $x_F$ using
the data sets of \cite{ref82}.  For comparison the NNLO AMP06 results
\cite{ref17} are shown as well.  To ease the graphical presentation
we have multiplied the results for the cross sections by $10^i$ with
$i$ indicated in parentheses in the figure for each fixed average value
of $x_F$.}
\end{center}
\end{figure}
\clearpage
\begin{figure}
\begin{center}
\ifpdf
\includegraphics[width=16.0cm]{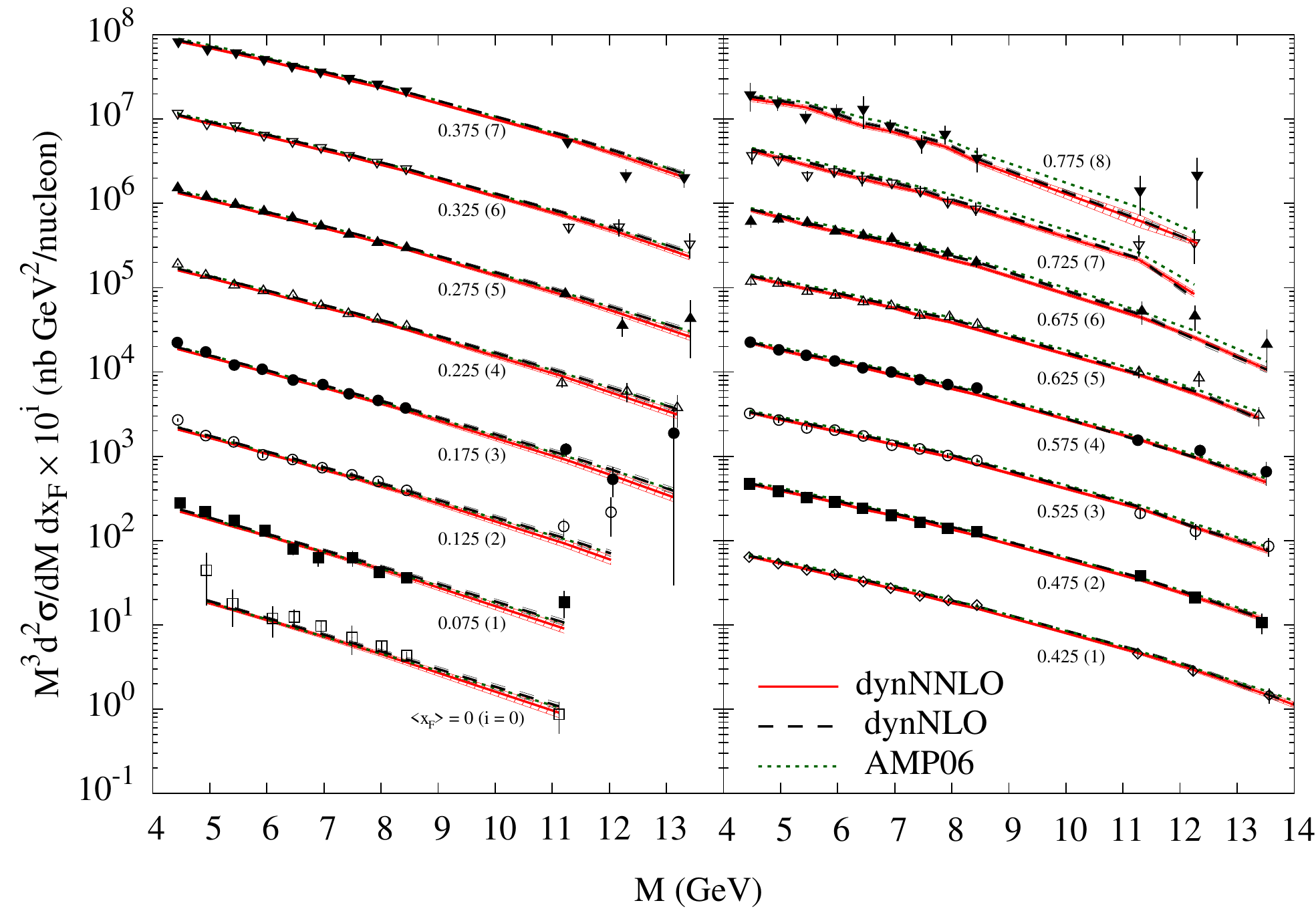}
\fi
\caption{As in Fig.~8 but for $pd$ collisions.}
\end{center}
\end{figure}
\clearpage
\begin{figure}
\begin{center}
\ifpdf
\includegraphics[width=11.0cm]{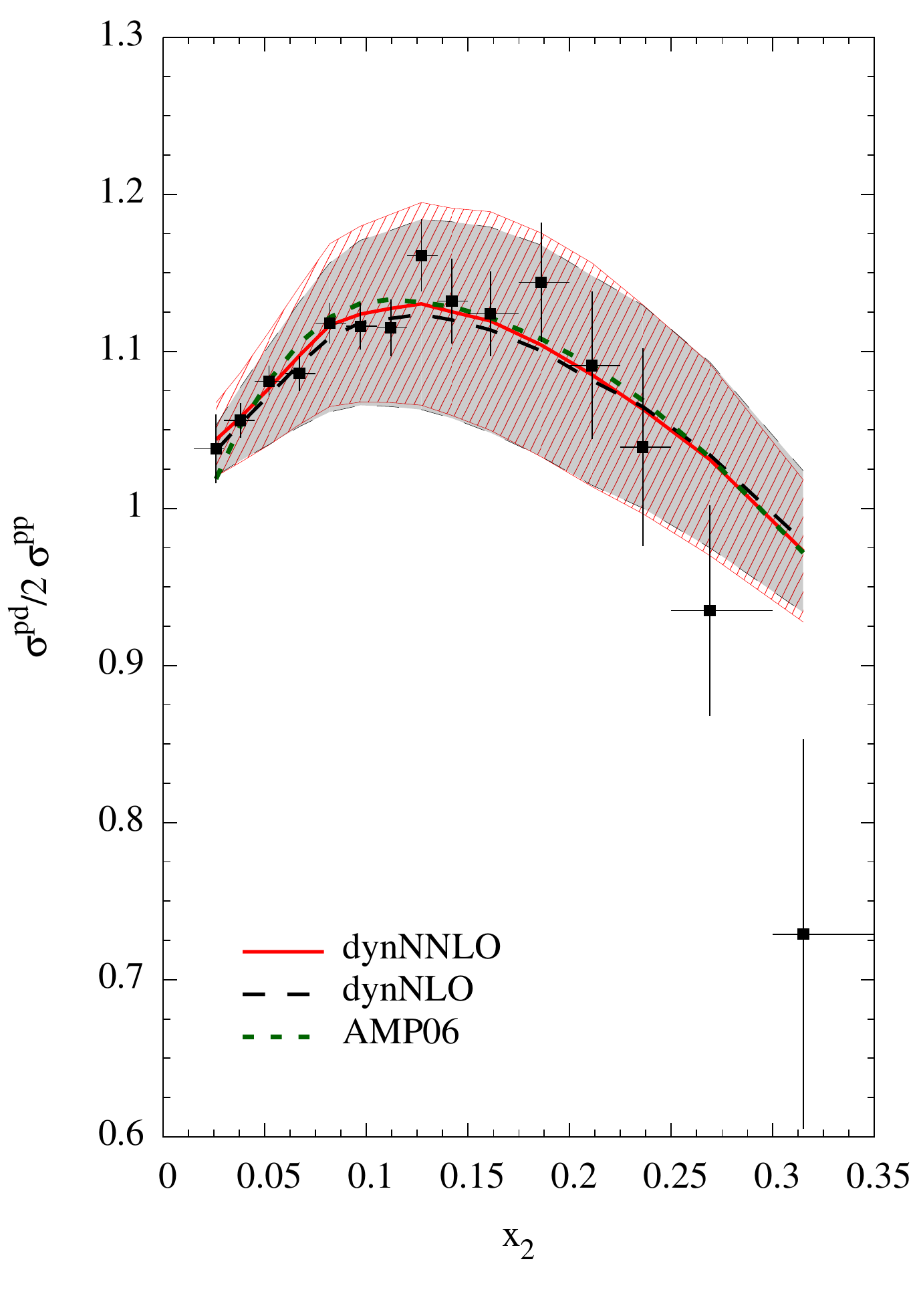}
\fi
\caption{Our dynamical NNLO and NLO results, together with their
$\pm 1\sigma$ uncertainties, for $\sigma^{pd}/2\sigma^{pp}$ as a 
function of the average fractional momentum $x_2$ of the target
partons.  The dynamical NLO results are taken from \cite{ref8}, and
the NNLO AMP06 ones from \cite{ref17}.  The data for the dimuon
mass range $4.6$ GeV $\leq M \leq 12.9$ GeV are from \cite{ref83}.}
\end{center}
\end{figure}
\clearpage
\begin{figure}
\begin{center}
\ifpdf
\includegraphics[width=14.0cm]{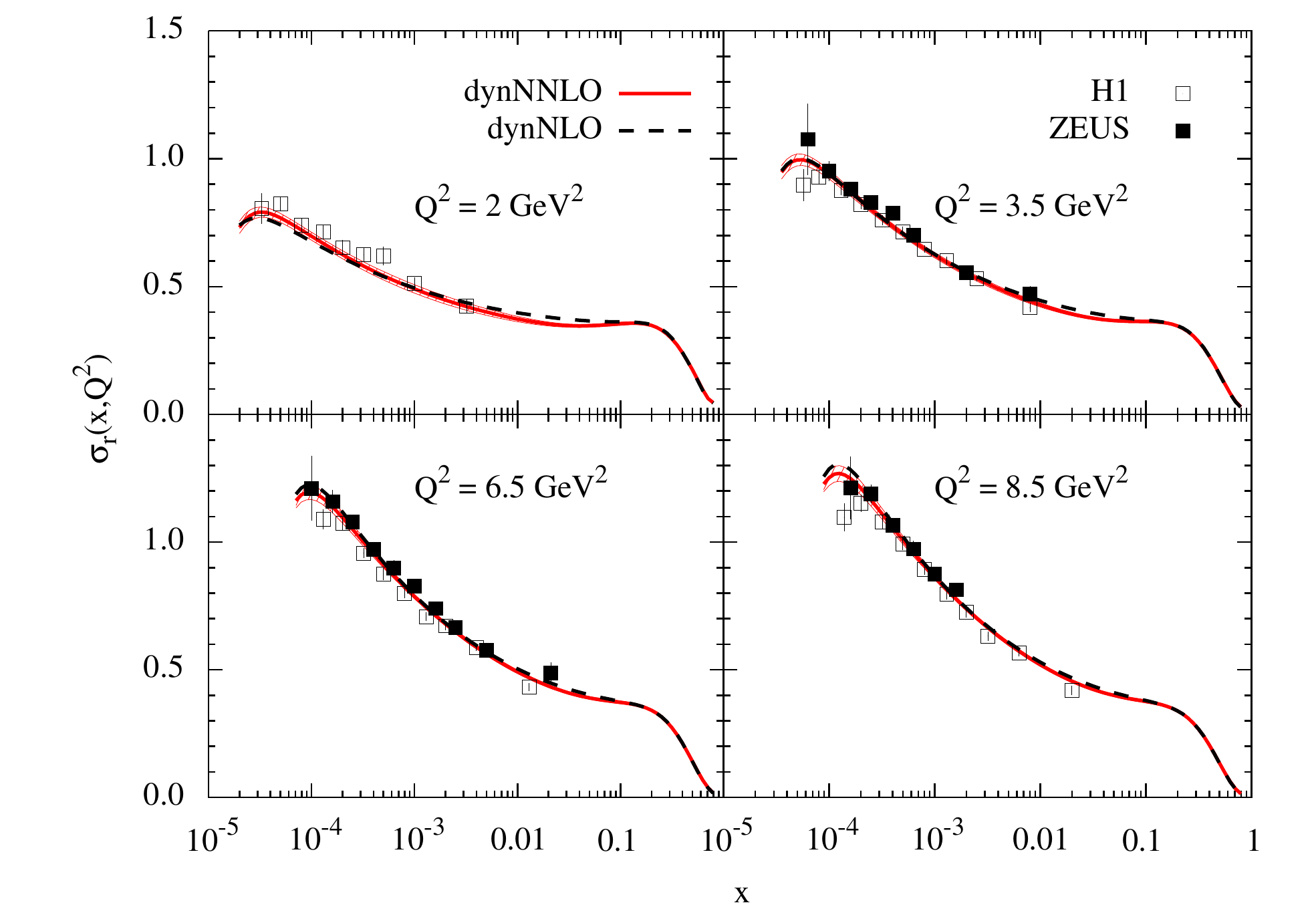}
\fi
\caption{The dynamical NNLO predictions, together with their 
$\pm 1\sigma$ uncertainties, for the  `reduced' DIS cross section
$\sigma_r(x,Q^2)=F_2-(y^2/Y_+)F_L$. The uncertainty bands of our
previous dynamical NLO results (dashed curves) are very similar
in size \cite{ref8} as the ones shown for NNLO.  The HERA data for
some representative fixed values of $Q^2$ are taken from [70--74].}
\end{center}
\end{figure}
\clearpage
\begin{figure}
\begin{center}
\ifpdf
\includegraphics[width=14.0cm]{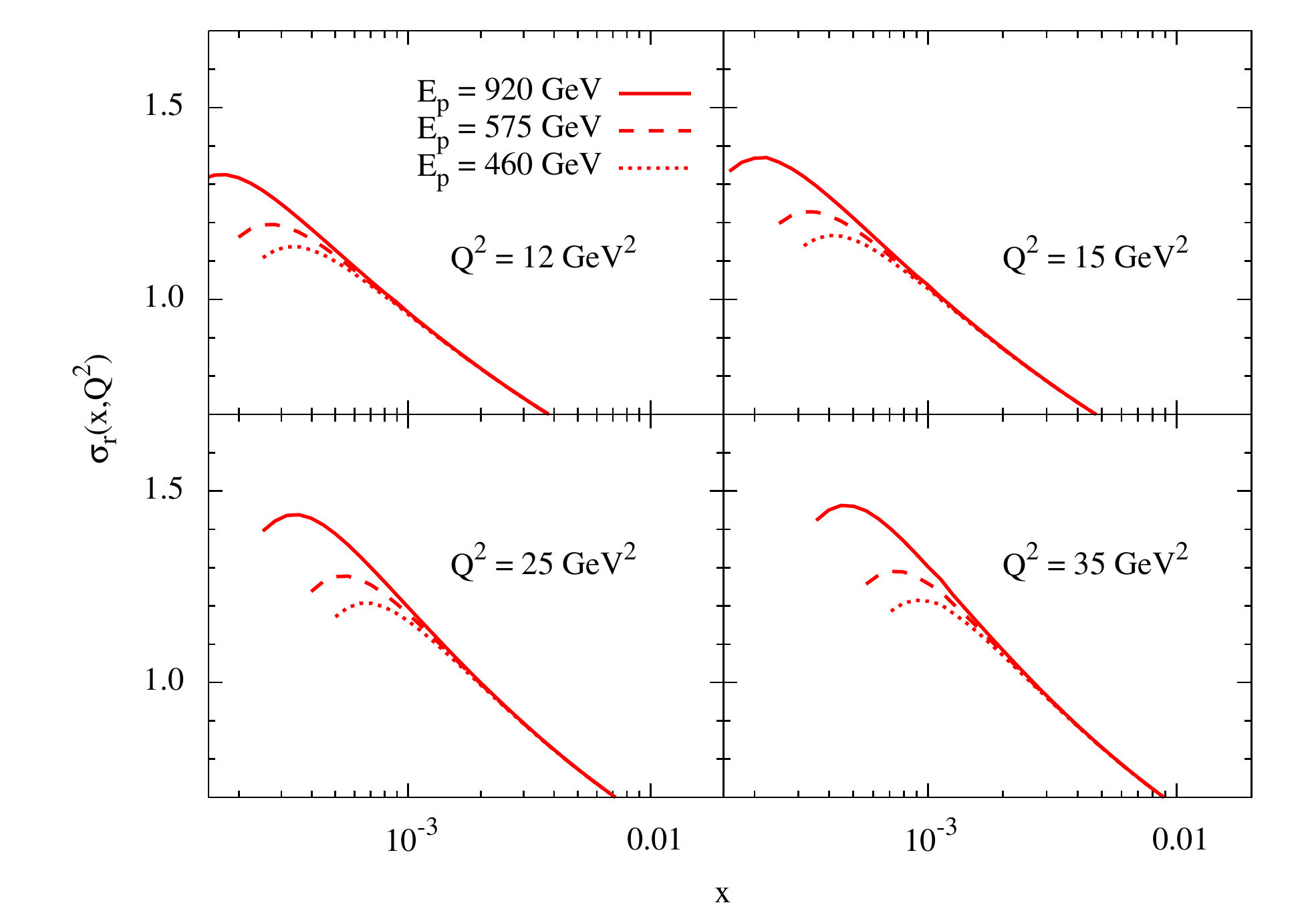}
\fi
\caption{Our dynamical NNLO predictions for $\sigma_r(x,Q^2)$ but for
different proton beam energies $E_p$ relevant for most recent HERA--H1
measurements \cite{ref96}. The $\pm 1\sigma$ uncertainty bands are 
similar to the ones shown in Fig.~11.  Notice that the curves
terminate when $y=1$.}
\end{center}
\end{figure}
\clearpage
\begin{figure}
\begin{center}
\ifpdf
\includegraphics[width=14.0cm]{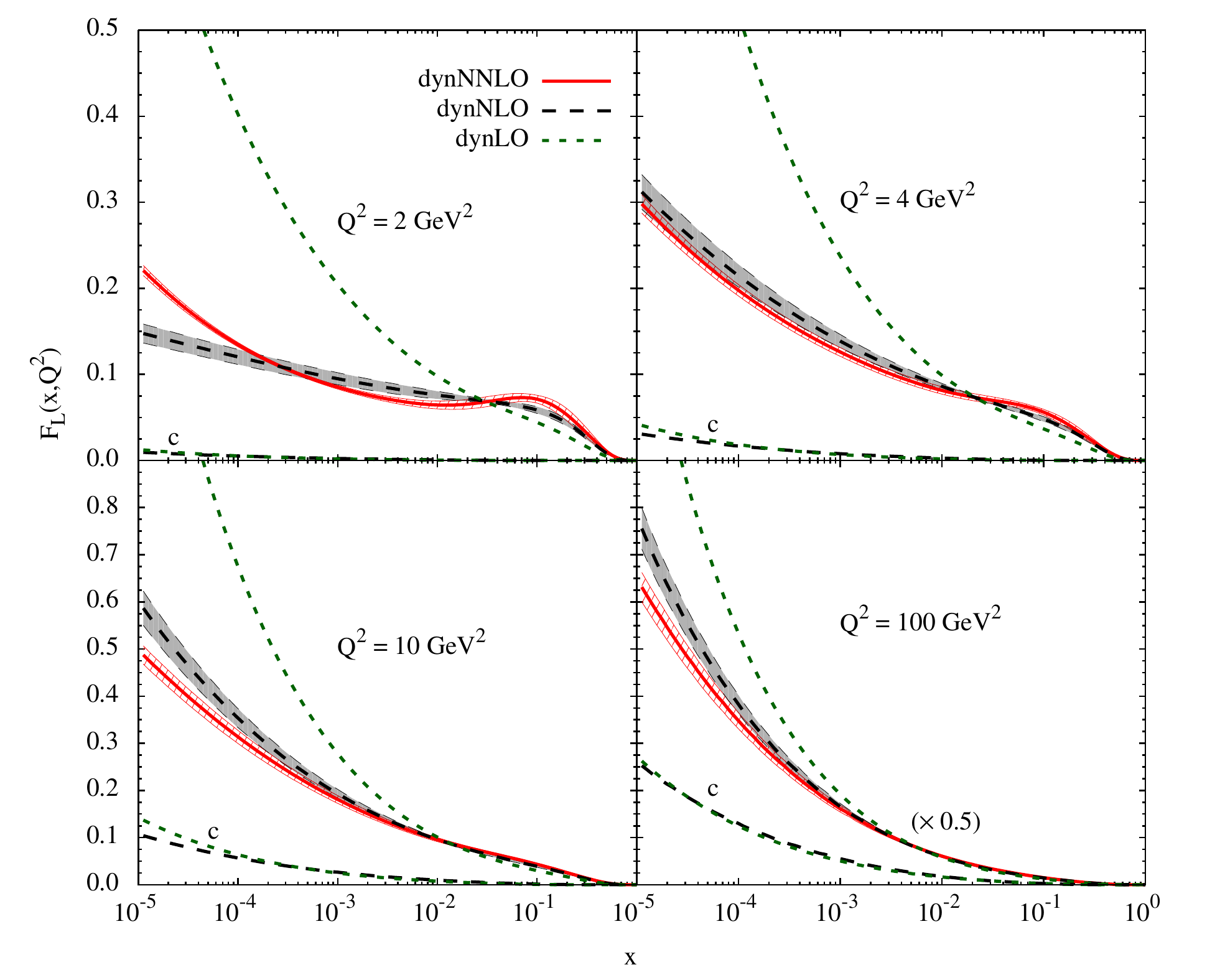}
\fi
\caption{Dynamical parton model LO, NLO and NNLO predictions for 
$F_L(x,Q^2)$ together with the $\pm 1\sigma$ uncertainty bands at NLO
and NNLO.  The heavy charm ($c$) contributions at LO (short--dashed
curves) and NLO (long--dashed curves) are shown as well.  The results
at $Q^2=100$ GeV$^2$ are multiplied by 0.5 as indicated.}
\end{center}
\end{figure}
\clearpage
\begin{figure}
\begin{center}
\ifpdf
\includegraphics[width=14.0cm]{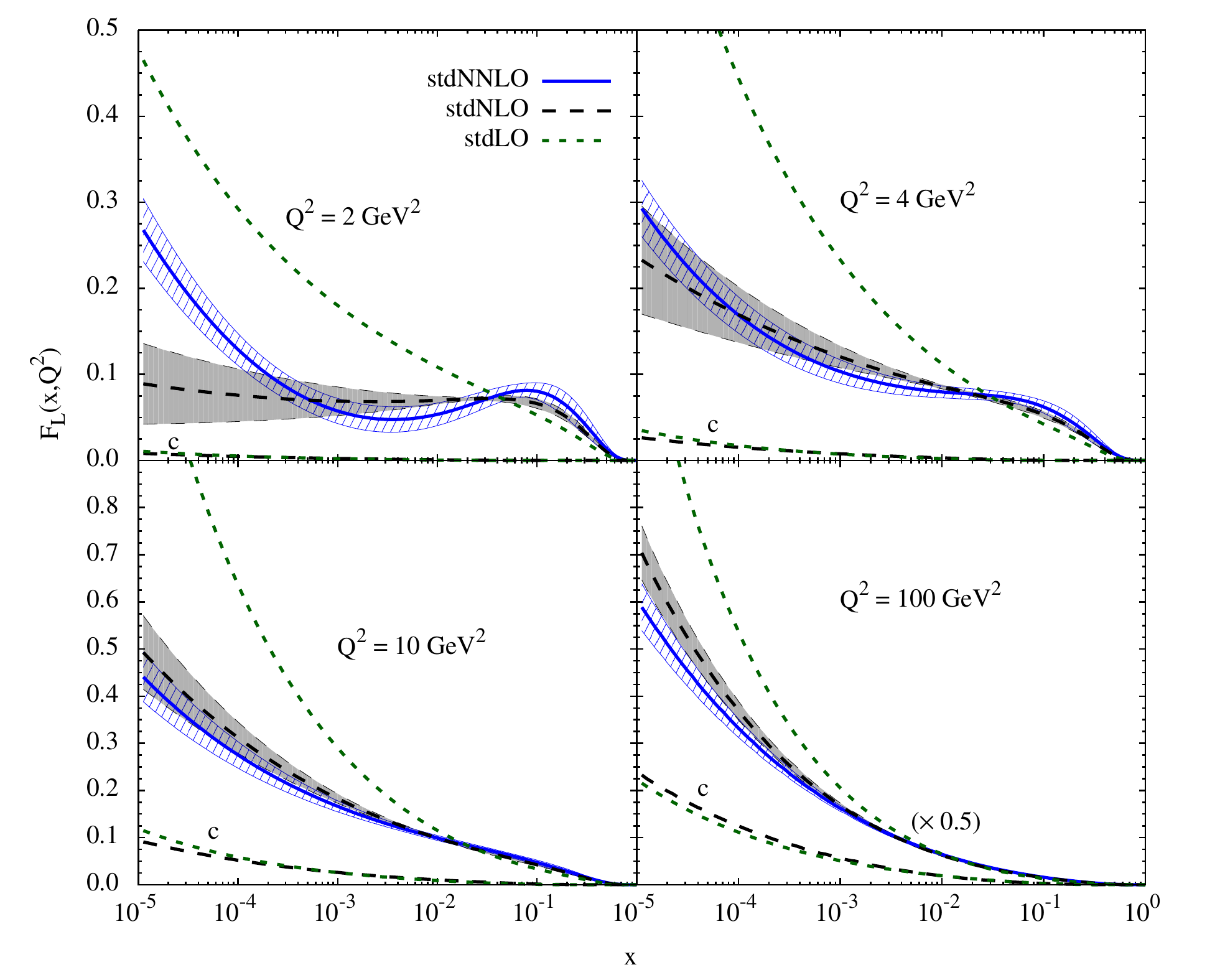}
\fi
\caption{As in Fig.~13 but for the common standard parton distributions.
Note that $Q^2=2$ GeV$^2$ coincides here with the input scale $Q_0^2$.}
\end{center}
\end{figure}
\clearpage
\begin{figure}
\begin{center}
\ifpdf
\includegraphics[width=14.0cm]{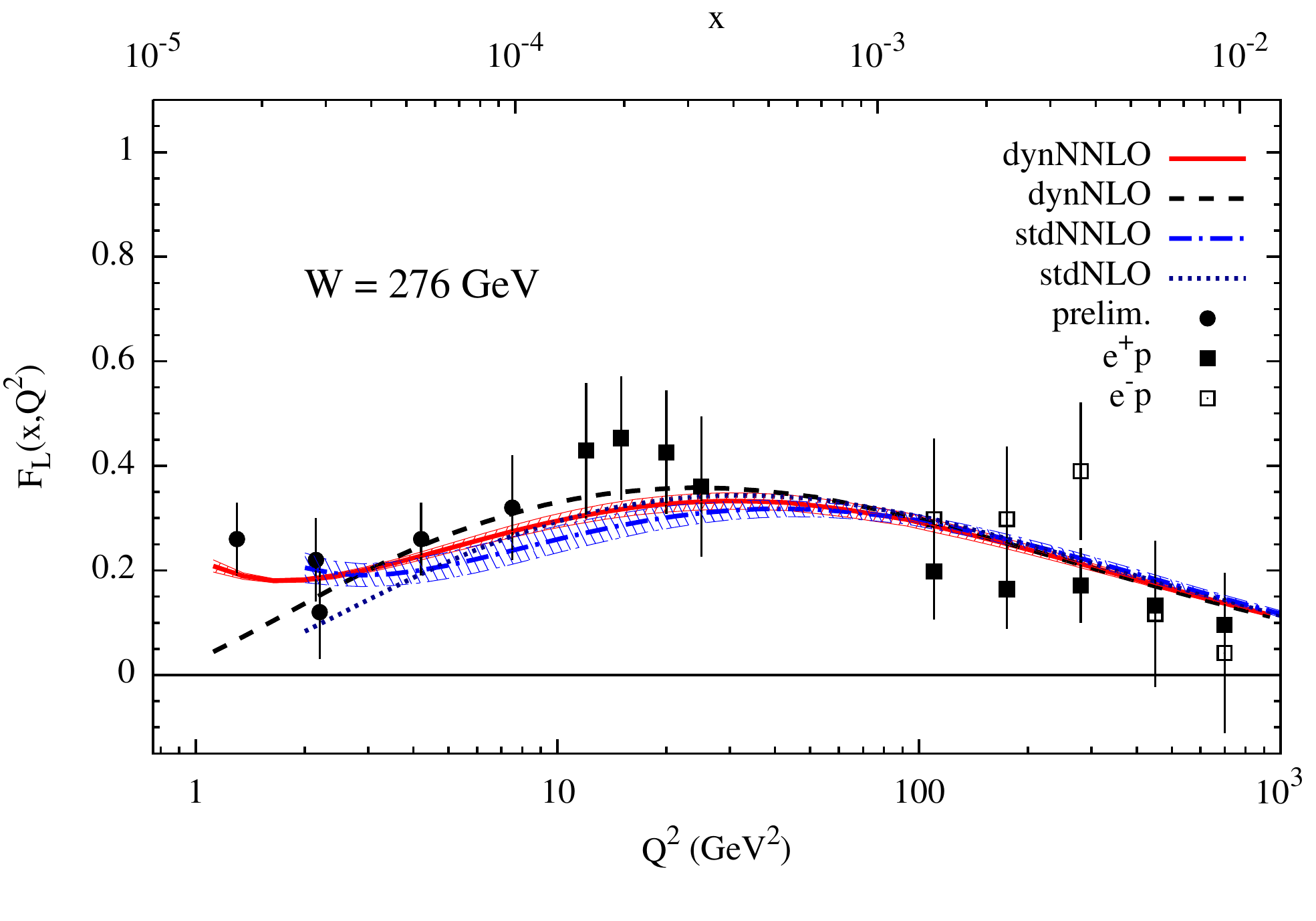}
\fi
\caption{Dynamical and standard NNLO and NLO predictions for $F_L(x,Q^2)$ 
at a fixed value of $W=276$ GeV.  The NLO($\overline{\rm MS}$) results
are taken from \cite{ref8}.  The (partly preliminary) H1 data
\cite{ref72,ref73,ref101,ref102} are at fixed $W\simeq 276$ GeV. The
more recent H1 data \cite{ref96}, which  correspond to smaller values of $W$
(larger $x$ and $Q^2$), are compatible with the indirectly
determined data shown.}
\end{center}
\end{figure}
\clearpage
\end{document}